\documentclass[a4paper,11pt]{article}

\font\scap=cmcsc10
\font\email=cmr10 at 8pt

\newcount\eqnumber
\def\neweq{{\rm{(\the\eqnumber)}}\global\advance\eqnumber by 1}
\def\eqdef#1{\eqno\xdef#1{\the\eqnumber}\neweq}
\def\newaeq{{\rm{(\the\eqnumber a)}}\global\advance\eqnumber by 1}
\def\eqdaf#1{\eqno\xdef#1{\the\eqnumber}\newaeq}
\def\newbeq{{\rm{(\the\eqnumber b)}}}
\def\eqdbf#1{\advance\eqnumber by -1\eqno\xdef#1{\the\eqnumber}\newbeq}
\def\newceq{{\rm{(\the\eqnumber c)}}}
\def\eqdcf#1{\advance\eqnumber by -1\eqno\xdef#1{\the\eqnumber}\newceq}

\def\eqdisp#1{\xdef#1{\the\eqnumber}\neweq}
\def\eqdasp#1{\xdef#1{\the\eqnumber}\newaeq}

\newcount\refnumber
\def\newref{{\the\refnumber}\global\advance\refnumber by 1}
\def\refdef#1{{\xdef#1{\the\refnumber}}\newref}

\newcount\fignumber
\def\newfig{{\the\fignumber}\global\advance\fignumber by 1}
\def\figdef#1{{\xdef#1{\the\fignumber}}\newfig}

\newcount\figappnumber
\def\newfigapp{{\the\figappnumber}\global\advance\figappnumber by 1}
\def\figappdef#1{{\xdef#1{\the\figappnumber}}\newfigapp}

\def\smallskip{\vskip 3pt}
\def\medskip{\vskip 6pt}
\def\bigskip{\vskip 12pt}

\usepackage{graphicx}
\usepackage[top=2.5cm,bottom=2.5cm,left=2cm,right=2cm]{geometry}

\begin{document}

\centerline{Modeling cooperation and competition in biological communities}
\bigskip
{\scap F. Meloni$^{1,2}$}, {\scap G. Nakamura$^{3,4,a}$},  {\scap B. Grammaticos$^{3,4}$},
{\scap A. S. Martinez$^{1,2}$}, and {\scap M. Badoual$^{3,4,b}$}

{\sl Faculdade de Filosofia, Ci\^{e}ncias e Letras de Ribeir\~{a}o Preto (FFCLRP), University of S\~{a}o Paulo, 14040-901 Ribeir\~{a}o Preto, Brazil}\phantom{ }$^{1}$

{\sl Istituto Nacional de Ci\^{e}ncia e Tecnologia em Sistemas Complexos (INCT-SC), 22290-180 Rio de Janeiro, Brazil}\phantom{ }$^{2}$

{\sl  Universit\'e Paris-Saclay, CNRS/IN2P3, IJCLab, 91405 Orsay, France}\phantom{ }$^{3}$

{\sl  Universit\'e de Paris, IJCLab, 91405 Orsay, France}\phantom{ }$^{4}$

{\email{gilberto.nakamura@ijclab.in2p3.fr}\phantom{ }$^{a}$

{mathilde.badoual@ijclab.in2p3.fr}\phantom{ }$^{b}$ }
\bigskip
{\sl Abstract}
\smallskip
The far-reaching consequences of ecological interactions in the
dynamics of biological communities remain an intriguing subject. For
decades, competition has been a cornerstone in ecological processes,
but mounting evidence shows that cooperation does also contribute to
the structure of biological communities. Here, we propose a simple
deterministic model for the study of the effects of facilitation and
competition in the dynamics of such systems. The simultaneous
inclusion of both effects produces rich dynamics and captures the
context-dependence observed in the formation of ecological
communities. The approach reproduces relevant aspects of primary and
secondary plant succession, the effect invasive species, and the
survival of rare species. The model also takes into account the role
of the ecological priority effect and stress the crucial role of
facilitation in conservation efforts and species coexistence.


\bigskip
1. {\scap Introduction}
\medskip

A major challenge in the theory of community organization is to
understand the origin of species diversity and its impact on the
structure of biological communities. The problem still persists to date mainly
because of the vast  number of abiotic and biotic factors that affect the
ability of species to invade and fix their habitat [\refdef{\Hoek2016},\refdef{\Begon2021}].
The abiotic factors include
environmental constraints which directly affect the demographics of
biological communities. The most notable examples are the spatial
effects, resource heterogeneity, environmental disturbances such as
vegetation fires or other natural phenomena, and events caused or
produced by climate changes [\refdef{\Hastings2007}]. In contrast, biotic
factors encompass an egregious number of ecological processes and
interactions among the various species that constitute the community.
The intricate network of inter- and intraspecific interactions
ultimately determines the ecological response of species at
population and community levels and shapes the formation and
stability of ecosystems [\refdef{\Ellner2018},\refdef{\Miele2019}].

Ecological interactions comprise the various types of influence that
one organism can exert over another. They are
subdivided into positive, neutral, or negative, depending on the
outcome onto the target organism. Interspecific predation and
competition are, perhaps, the most well-known types of interactions:
they embody the natural dispute for resources in ecology and set the
cornerstones of the mathematical models developed by Volterra [\refdef{\volterra}],
MacArthur [\refdef{\mcarthur}], and Levin [\refdef{\levin}] to name a few. Together with Gause's
principle of species exclusion [\refdef{\Gause1935}] -- species competing
for the same resource cannot coexist -- and Hutchinsons' niche
separation [\refdef{\Hutchinson1957}], their works established the classical
theory of ecological communities. In this context, ecologists have
investigated the response of ecosystems to species invasion and
exclusion, stress, environmental changes, and the impact on the
structure and functionality of ecosystems
[\refdef{\Fischer2019},\refdef{\Clark2019},\refdef{\Prach2019},\refdef{\Changb2019}].

However, the existence of widespread species diversity observed across
the globe poses a conceptual challenge to the classical theory. More
specifically, to the principle of species exclusion. The empirical
observation that different marine species competing for the same set
of resources can coexist summarizes what is known as the plankton
paradox [\refdef{\Hutchinsonb1961}]. According to the classical theory, and
assuming homogeneous conditions, only the most competitive species
should survive. It turns out that violations of species exclusion
occur fairly easily in nature. Palmer lists 120 ecological processes
that violate the principle and lead to species coexistence
[\refdef{\Palmer1994}]. In that regard, the modern theory of coexistence by 
Chesson [\refdef{\Chesson2000}] synthesize many of those processes. It
states that the suppression of the growth rate is stronger for higher
intraspecific densities rather than interspecific ones. This gives a small
advantage to invasive species at low densities and promotes
coexistence via the decrease of the effective growth rate of the
dominant species with their population sizes.

The inability to accommodate small deviations to common assumptions
suggests the classical theory still requires adjustments. In
particular, the role of positive interactions is generally neglected 
even though several ecosystems have evolved and  still rely on positive
interactions. Coral reefs are prime examples. The mutualism between
corals and algae (dinoflagellates) allows them to thrive in sub-optimal
environmental conditions [\refdef{\Yuyama2018}]. Sediments and thermal
stress quickly disrupt this delicate interaction, leading to the
removal of algae and, eventually, to the characteristic bleaching
of coral reefs [\refdef{\Fisher2019}]. This peculiar phenomenon -- and
linked with climate changes -- highlights the dependence of positive
interactions in the maintenance of complex ecosystems.

Facilitation can also be an important tool for invasive species, a
fact often overlooked in competition-driven models. Based on empirical
data, the research team in Ref.~[\refdef{\Moyano2020}] found out that
seedling strategies for highly invasive trees of the genus {\it Pinus}
hinge on the mutual relation with mycorrhizal fungi. The positive
interaction with the fungi improves the transfer of nutrients and
minerals from the soil to the seedling, promoting its growth and
increasing the chances of survival. As a result, the invasive trees
tend to produce lighter seeds in greater numbers to overcome the
reduced chance of finding the fungi. In contrast, the authors of Ref.~[\Moyano2020] show
that less invasive trees produce seeds that are less dependent on
mycorrhizal fungi and thus larger.

The inclusion of facilitation in a framework driven by competitive
interactions has proven to be a challenging task so far
[\refdef{\Gross2008}]. A cascade of facilitation processes is known to
occur in plant successions at empirical and conceptual levels
[\refdef{\Tilman1990},\refdef{\Muller2018}]. Plants compete directly for nutrients,
water, and sunlight. Early species are more tolerant to poor
distribution of nutrients in the soil and, as a result, they facilitate
the germination of late-stage plants by either increasing water
content or locally reducing the temperature of soil [\refdef{\Bruno2003}]. 
This positive interaction benefits late species which eventually
suppress or replace the early species as they are more efficient to
capture sunlight and other resources. The repetition of this process
eventually leads to plant or forest succession and the modification of
the ecosystem. The key lesson here is that large densities of early
species affect the dynamics of late species in different manners
depending on their development phase. The behavioral changes figure among
one of the factors that complicate a broader formulation of community
structures.  

Furthermore, the introduction of facilitation processes modifies the
basic tenets of niche theory. As discussed in detail [\refdef{\Brunob2016}],
interactions that promote facilitation between plants also create
paths for species coexistence and increased species richness. The
current consensus claims that diversity reduces the effects of
abiotic stresses and helps to maintain the stability of ecosystems. At
the same time, a larger number of species also implies an increased
likelihood of species competing for the same niche, clashing again
with Gause's exclusion principle. The solution for the apparent
paradox is that facilitation processes can expand niches while
competition reduces them. The expansion of niches can be obtained
by merely increasing the output of an existing niche; or by
allowing organisms to access species-transformed or mediated
resources. The latter actively shape the distribution and
accessibility of resources in ecosystems instead of merely control
their amplitudes (competition). Thus, the inclusion of facilitation
forces the departure from the pragmatic view of niche as the
representation of {  locales} where species can exist.

Here, we propose a simple dynamical model of plant communities
that explicitly contains interspecific competition and facilitation.
The inclusion of positive interactions creates a rich dynamics that
can reproduce the basic aspects of forest succession, species
coexistence, and response to catastrophic events.
Our findings show that our model can reproduce the general
aspects of experiments testing the stress gradient hypothesis, a
phenomenon in which the behavior of interacting species can vary from
pure cooperative to pure competition according to their stress level.
The paper is organized as follows. In Sec.~2, we revisit some models of
population dynamics and the manner in which they approach the
interaction between species. Inspired by the Kuno model, in Sec.~3, we
extend the formulation to include facilitation as a core part of the
dynamics and explain the discrete model for two and three species. The
results are shown in Sec.~4 and further discussed in Sec.~5. Our
concluding remarks are offered in Sec.~6.


\bigskip
{\scap Box 1: Ecological succession}
\medskip

Ecological succession is the natural process that describes the
assembling and evolution of species structures in ecological
communities. At its core, it describes how the organisms gradually
modify their surrounding and create conditions for the establishment
of species or their eradication. The process often yields a global
increment in biomass and changes the composition of species. The
characteristics and ecological functioning of the area 
also change {  in order} to reflect the new set of dominant species. 
Forest succession is the prime example: plants grow in an ordered and
predictable pattern marked by the progressive dominance of long-living
species. In a nutshell, from bare soil to grass, to shrubs and small
trees, culminating in {  mature} trees (see Figure~{\figdef{\figbox}}).    
In the early stages, pioneer species are dominant under harsh
conditions and actively alter the physico-chemical properties of soil
and surrounding areas. The presence of pioneers {  ensures} adequate
conditions for secondary species to {  get a foothold} and grow. Although
secondary species are less adapted to harsh conditions, they exhibit
better competitive traits and, thus, tend to replace pioneers,
completely or partially. The successional process continues and
eventually reaches the climax stage (steady state).

Successional dynamics are classified as primary or secondary. Primary
succession concerns the initial colonization of ecosystems. It starts
from soil formation and proceeds all the way up to the climax stage.
Secondary successions occur shortly after ecosystems are disturbed by
natural or human-made events and return
to a previous developmental stage. The main difference between
secondary and primary succession is the initial distribution of
species, with reduced or complete absence of earlier species depending
on the intensity and type of disturbance. According to the classical
concept by Clements [\refdef{\Clements1916}], the ecosystem should return to the same steady
state prior to the disturbance. However, mounting evidence 
suggests that secondary successions may lead to different late
successional stages, with significant variation of species
composition [\refdef{\Cramer2008}, \refdef{\Estrada2020}]. Possible
explanations include reduced biomass and species diversity caused by
phosphorus depletion [\refdef{\Peltzer2010}, {\Muller2018}], and spatial or
priority effects during the colonization [\refdef{\Damschen2008},
  \refdef{\Meloni2015}, \refdef{\Demeester2016}, \refdef{\Fukami2015}]. These hypotheses indicate
that initial conditions are important drivers for late
stages in ecological succession [\refdef{\Chang2019}].

\medskip
\centerline{\includegraphics[width=10cm,keepaspectratio]{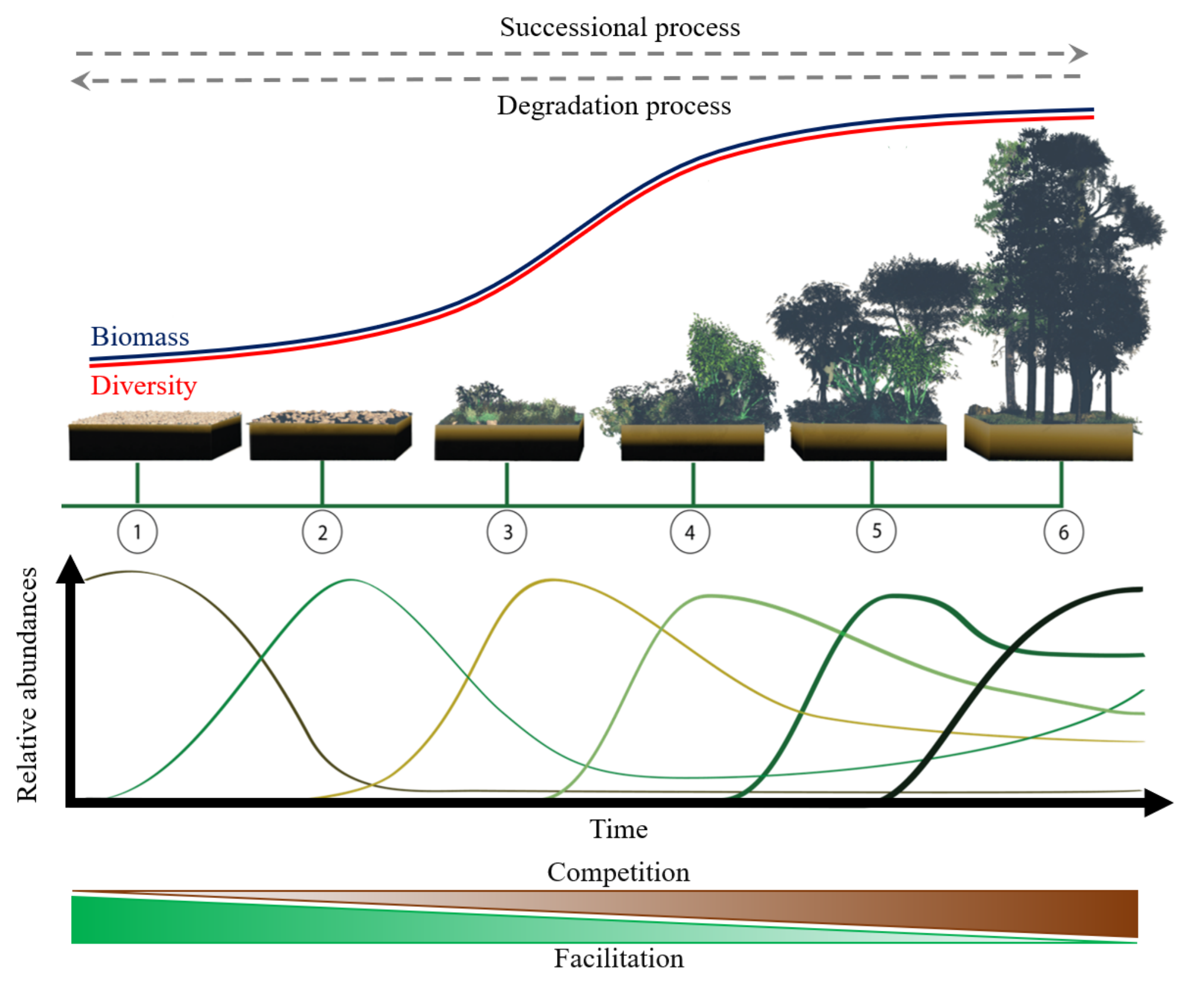}}
Figure~{\figbox}: Schematic representation of ecological succession dynamics of a forest ecosystem. The changes in relative abundances indicates the shifts in the structures of biological communities. Facilitation dominates early stages while competition becomes the dominant interaction between species at late stages. Modified from Lucas Martin Frey 2011 (Creative Commons Attribution 3.0.). 
\medskip

\bigskip
{\scap Box 2: Biological Invasion}
\medskip

The colonization of new areas by organisms is a fundamental aspect of
ecology. Colonization contributes to the formation and maintenance of
ecosystems, being crucial for the dynamics of natural communities. For
human-made systems, such as plantations, biological invasions can
negatively affect the ecological balance and food production. Recent
estimates place the annual cost of losses due to biological invasions
around US\$ 27 billion, reaching a stunning mark of US\$ 1.3 trillion
over the course of 47 years [\refdef{\Diagne2021}]. This outrageous sum and
the dire consequences to food security {  highlight} the necessity to
understand how biological invasions happens and what influences them.

In broader terms, the successful colonization depends on the species'
ability to surpass the ecological filters of the target region.
Filters are the collection of environmental and ecological
aspects that in someway suppress or restrict the survival of
organisms in a given habitat. They can represent harsh conditions like
dryness, temperature, presence of predators, etc.
Strong ecological filters hinder the development of species so much
that only a handful of specialized species can flourish. These species
tend to take part in facilitation processes with others in order to
counterbalance the stress imposed by the environment. 
In contrast, {weak filters provide natural shelters} and
foster a far more diverse, but less specialized, biological community
characterized by ample competition between species. The amplitudes of
these competitions constitute yet another ecological filter, a
dynamical one, with a stronger dependence on the current state of the
ecosystem.

In the context of colonization, dynamical filters are known as
ecological priority effects and they can influence the organization of
biological communities [\refdef{\Demeester2016}]. For instance, early
colonizers can monopolize resources and maintain their dominant status
even against superior competitors due to their sheer abundance
(founder effect), inverting the usual species exclusion. These effects
must be taken into account by restoration and conservation efforts,
especially those that aim to preserve or reintroduce native species to
target locations.

\bigskip
2. {\scap Models of population dynamics}
\medskip
The question of human population growth has been one of vital importance since the dawn of civilisation. Regular censuses allowed to keep track of the evolution and the compilation of their data were a precious aid in the formulation of modelling-based descriptions of the dynamics. Given the reproductive mechanisms, it is clear that an initial population, if left unchecked, will grow in a geometric way. Malthus [\refdef\malthus] was the first in addressing the question in a rigorous way, pointing out that the population growth cannot be indefinitely sustained, given that the resources do not increase with the same rate. The geometric growth to which Malthus refers to is what in modern parlance we call an exponential increase, governed by the equation
$${dN\over dt}=\lambda N\eqdef\zena$$
where $N$ is the population. The solution of (\zena) is simply $N=N_0e^{\lambda t}$. 

The fact that the scarcity of resources can curb the exponential growth of the population, was incorporated in a mathematical formulation proposed by Verhulst [\refdef\verhulst]. The equation, known today under the moniker of logistic, has the form
$${dN\over dt}=\lambda \left(N-{N^2\over K}\right),\eqdef\zdyo$$
with solution
$$N={N_0K\over N_0+(K-N_0)e^{-\lambda t}}.\eqdef\ztri$$
The $N^2$ term, which is essential for halting the exponential growth, is what we would call a self-interaction term, encoding the interspecies competition for resources. When $t\to\infty$ the population tends to the value $K$, which is often referred to as the carrying capacity of the environment, corresponding to the maximal value of the population that can be sustained. We shall not pursue the discussion of the Verhulst equation, often criticised for its extreme simplicity, but we shall present, as an aside, the work of Morisita [\refdef\morisita], who addressed the question of the form of the logistic equation when one considers that the population grows by spurts occurring at regular time intervals. Morisita obtained the recursion relation
$$N_{t+\tau}-N_t=\mu\left(N_t-{N_tN_{t+\tau}\over K}\right),\eqdef\ztes$$
where $\tau$ is the time step and $\mu=e^{\lambda t}-1$. It turns out that (\ztes) is satisfied by (\ztri) when the latter is sampled over the grid points $t_0+n\tau$. In the next section we shall return on the Morisita approach, with further comments.

The situation becomes more interesting when one considers two
interdependent species. The classical situation is the one known as a
predator-prey one, where one species (the predator) depends on the
other (the prey) for its subsistence. The simplest mathematical
formulation of such a situation is that proposed by Lotka
[\refdef\lotka] and, independently, by Volterra [\volterra]. The proposed model uses the chemical principle of mass action. Namely, the interaction between the populations is represented by the product of their biomass densities. The system thus assumes the form
$${dn\over dt}=\lambda n-anp\eqdaf\zpen$$
$${dp\over dt}=bnp-\mu p.\eqdbf\zpen$$
One remarks that the predator $p$ cannot survive in the absence of prey, while the population of the latter, in the absence of predator, grows unchecked. In this respect it would have been more natural to introduce a slightly modified system where the population of the prey is governed by a logistic equation
$${dn\over dt}=\lambda n\left(1-{n\over k}\right)-anp\eqdaf\zhex$$
$${dp\over dt}=bnp-\mu p.\eqdbf\zhex$$
A major contribution to the theory of predator-prey systems was the introduction of what Salomon [\refdef\salomon] called the functional response. He argued that, since a predator can handle only a finite number of prey per unit time, the death rate of the the prey, the $-anp$ in (\zhex a), must depend non-linearly on the prey density, i.e. $-a(n)p$. Holling [\refdef\holling] proposed an explicit realisation of these ideas by introducing, what is known today as, the disk equation. According to him a suitable expression for $a(n)$ is
$$a(n)={fn\over n+\kappa},\eqdef\zhep$$
which happens to be identical to the Michaelis-Menten  [\refdef\menten] equation for enzyme kinetics. Combining all these features one of the authors (BG) introduced in [\refdef\crypto] the predator-prey model
$${dn\over dt}=\lambda n\left(1-{n\over k}\right)-ap{n\over1+\sigma n}\eqdaf\zoct$$
$${dp\over dt}=bp{n\over1+\rho n}-\mu p.\eqdbf\zoct$$
Having discussed the classic population dynamics models, we turn now
to the ones pertaining directly to the theme of this paper. While we
shall be interested in species which are in direct competition for
resources we shall eschew the case where upon species acts like a resource for the other. A successful model describing the situation of competition through reproductive interference is the one due to Kuno. In  [\refdef\kuno] he proposed a model involving two competing species which we can rewrite by reorganising the terms with respect to the original formulation (for reasons that will become clearer later) as
$$x'=a\left({x\over x+\alpha y}-fx\right)x-\lambda x\eqdaf\zenn$$
$$y'=b\left({y\over y+\beta x}-gy\right)y-\mu y,\eqdbf\zenn$$
where $x$ and $y$ are the biomass densities of the two species and the prime indicates a derivation with respect to time. Here, the interference between the species affects the birth rate of each as a diluting effect. Kuno presents, in parallel to this model, a Lotka-Volterra one where the two species compete for a common resource
$$x'=c\left(1-fx\right)x-\alpha xy\eqdaf\zdek$$
$$y'=d\left(1-gy\right)y-\beta yx.\eqdbf\zdek$$
Note that (\zdek) does {\sl not} describe a predator-prey system, the interaction term being noxious to both species, each of which would have survived and grown to capacity in the absence of the other. We shall return to the Kuno model in the next section. 
\bigskip
3. {\scap Building a model for competing and cooperating species}
\medskip
Before considering the case of two competing species, we shall revisit the case of a single species investigating possible extensions of the simple logistic equation. Our approach will be, in some remote sense, inspired by the work of Aubier who introduced a density dependent mortality term in a model aiming at describing an asymmetric competition between two species. Here we shall ignore the two-species case focusing on the density dependent mortality effect on a single species . The corresponding equation has the form
$$x'=ax\left(1-{x\over k}\right)-\lambda{x\over 1+\sigma x}\eqdef\dena$$
We now try to bring (\dena) as close as possible to the form of Kuno equation. We obtain
$$x'=ax\left({cx+d\over 1+\sigma x}-{x\over k}\right)-\lambda x\eqdef\ddyo$$
Note that when all coefficients in (\dena) are positive, the $c$ in (\ddyo) is negative. This is a specific feature of the model of Aubier and need not concern us further. The conclusion we can draw from (\ddyo) is that an extension of the logistic model is possible where the term controlling the capacity is a homographic function of the density. This leads thus to the following expression, obtained after the appropriate scaling of the $x$ variable
$$x'=ax\left({ x+\beta\over x+\gamma}-x\right)-\lambda x\eqdef\dtri$$
We claim that the solution of (\dtri) behaves exactly like the solution of the logistic equation. Clearly when $\beta=\gamma$ (\dtri) is just a way of rewriting (\zdyo), but the overall behaviour is the same even when $\beta,\gamma$ are unrelated. 
Before proceeding to a numerical simulation of (\dtri) in order to confirm our claim we must produce a discrete analogue thereof. For this we shall use the method one of the authors (BG) developed in [\refdef\ramani]. It is a method directly inspired from the works of Mickens [\refdef\mickens]. The best way to describe this discretisation procedure is by using the catchy phrase published in [\refdef\handy]: `if all quantities are positive, no minus sign should appear anywhere'. In the present situation we deal indeed with quantities that are by principle positive, being biomass densities and thus the prescription just cited can apply. The way to implement this is through the appropriate staggering of the variables. In order to show how the method works we shall revisited the approach of Morisita. We start from the logistic equation
$$x'=\lambda x-\mu x^2,\eqdef\dtes$$
introduce a forward difference for the derivative and rewrite the $x^2$ as $x_nx_{n+1}$ instead of $x_n^2$. We have thus ($\delta$ being the discretisation step):
$${x_{n+1}-x_n\over \delta}=\lambda x_n-\mu x_nx_{n+1}.\eqdef\dpen$$
This leads to the mapping
$$x_{n+1}=x_n{1+\lambda \delta\over1+\mu \delta x_n}\eqdef\dhex$$
which is precisely the one we obtain from the recursion of Morisita, upon solving for $N_{t+\tau}$.
We turn now to (\dtri) and again introduce a forward difference for the derivative and obtain:
$${x_{n+1}-x_n\over \delta}=ax_n{ x_n+\beta\over x_n+\gamma}-ax_nx_{n+1}-\lambda x_{n+1}\eqdef\dhep$$
leading to the mapping
$$x_{n+1}=x_n{1+a\delta{ x_n+\beta\over x_n+\gamma}\over1+\lambda \delta+a\delta x_n},\eqdef\doctp$$
which can be used to integrate (\dtri). 
\medskip
\centerline{\includegraphics[width=10cm,keepaspectratio]{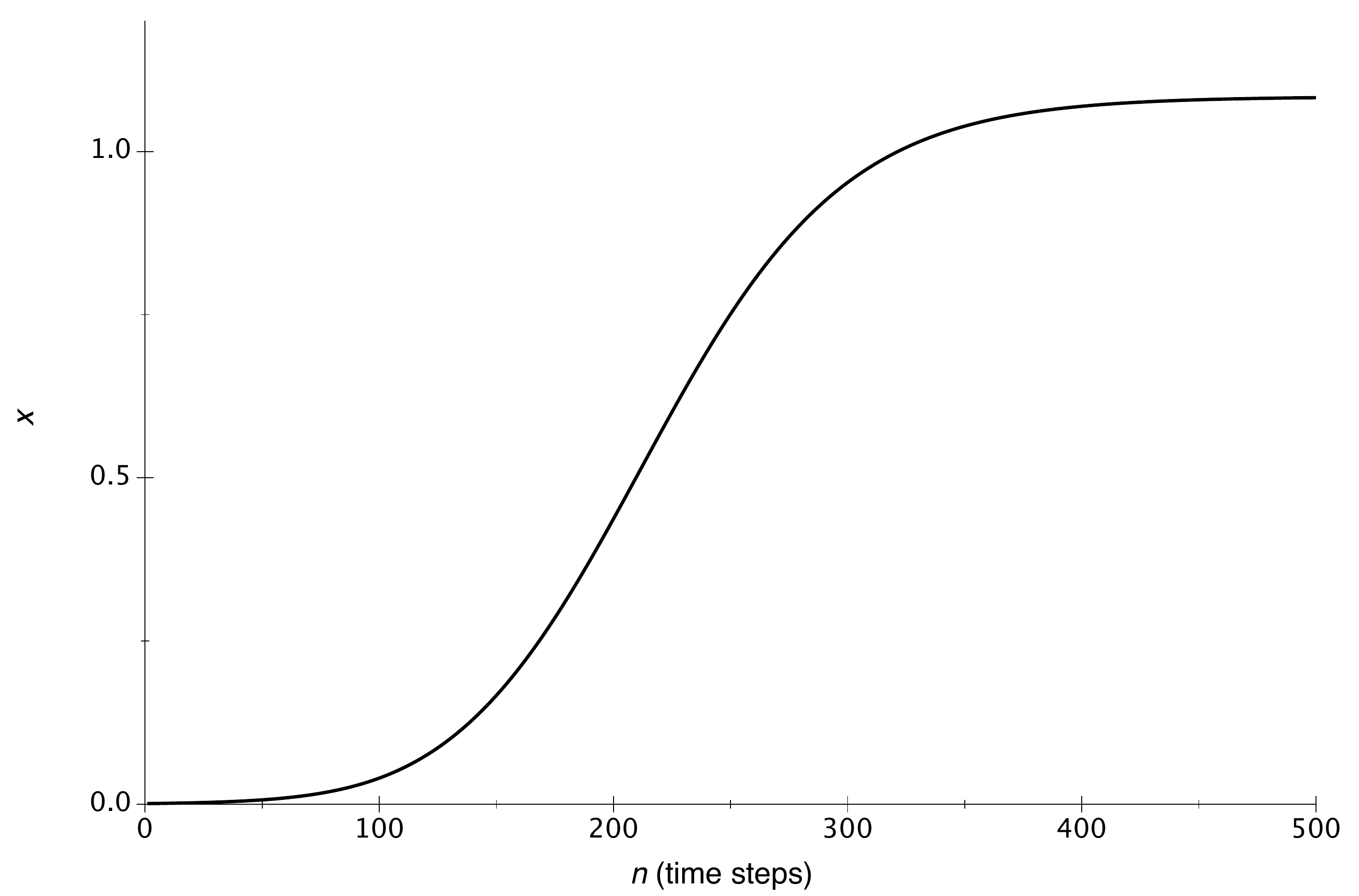}}
\smallskip{\bf Figure \figdef\one}. {\it Evolution of the biomass $x$ versus time for model (\dtri). Parameters: $x(0)=1~10^{-3}$, $a=2.0$, $\beta=0.2$, $\gamma=0.1$, $\lambda=10^{-3}$}
\medskip
Figure \one~shows such a simulation and the resemblance to the solution of the logistic is uncanny. Note that the asymptotic value of $x$, which plays the role of the ecological capacity, is obtained in a complicated way from the parameters of the model. However, and this is an important feature of our discretisation approach, the fixed points and their stability conditions are exactly the same in the continuous and the discrete system.

Having set the model for a single species we turn now to the case where two species are present. The difference of our approach and that of Kuno is that we assume that the two species may not only compete but also cooperate. 
At this point a remark is in order. The terms `cooperation' and `competition' are used rather loosely. In the present context cooperation should better be understood as `facilitation' whereupon a species profits from the presence of the other. In the same spirit competition is to be understood as `hindrance', one species preventing the other to realise its full potential. Having clarified the real meaning of the terms we shall be using throughout the paper, we can proceed to the construction of our model.

Starting from the equations (\zenn) where only competition is present we extend equation (\dtri) introducing a capacity involving also cooperation terms leading to the system
$$x'=a\left({ x+f_{xy} y\over x+h_{xy} y}-x\right)x-\lambda x\eqdaf\denn$$
$$y'=b\left({ y+f_{yx} x\over y+h_{yx} x}-y\right)y-\mu y.\eqdbf\denn$$

where the terms $f_{xy} y$ (repectively $h_{xy} y$) quantifies how much the growth of species $x$ is facilitated (resp. hindered) by the presence of  species $y$. Clearly the case of pure, Kuno-like, competition can be obtained from (\denn) by putting $f_{xy}=f_{yx}=0$. A pure cooperation situation would correspond to $h_{xy}=h_{yx}=0$. 

Before proceeding further it is interesting to study the fixed points of (\denn) as well as their stability properties. Three distinct situations do exist: both fixed points are 0, one fixed point is 0 while the other is finite and finally both fixed points can be finite. 
The stability of the fixed points (0,0) leads to the consideration of the system
$$\xi'=\Lambda \xi-\lambda \xi\eqdaf\ddyo$$
$$\psi'=M \psi-\mu\psi\eqdbf\ddyo$$
where $\Lambda$ and $M$ are the limits of the expressions $a( x+f_{xy} y)/(x+h_{xy} y)$ and $b( y+f_{yx} x)/(y+h_{yx} x)$ when $x\to0$ and $y\to0$. However the values of $\Lambda$ and $M$ depend on the precise way $x$ and $y$ tend to 0 and thus it is not easy to give a criterion for the attraction to the point $(0,0)$. However it is clear that when the parameters $\lambda$ and $\mu$ are large enough the solution of (\denn) is attracted to $(0,0)$. 
\medskip
\centerline{\includegraphics[width=10cm,keepaspectratio]{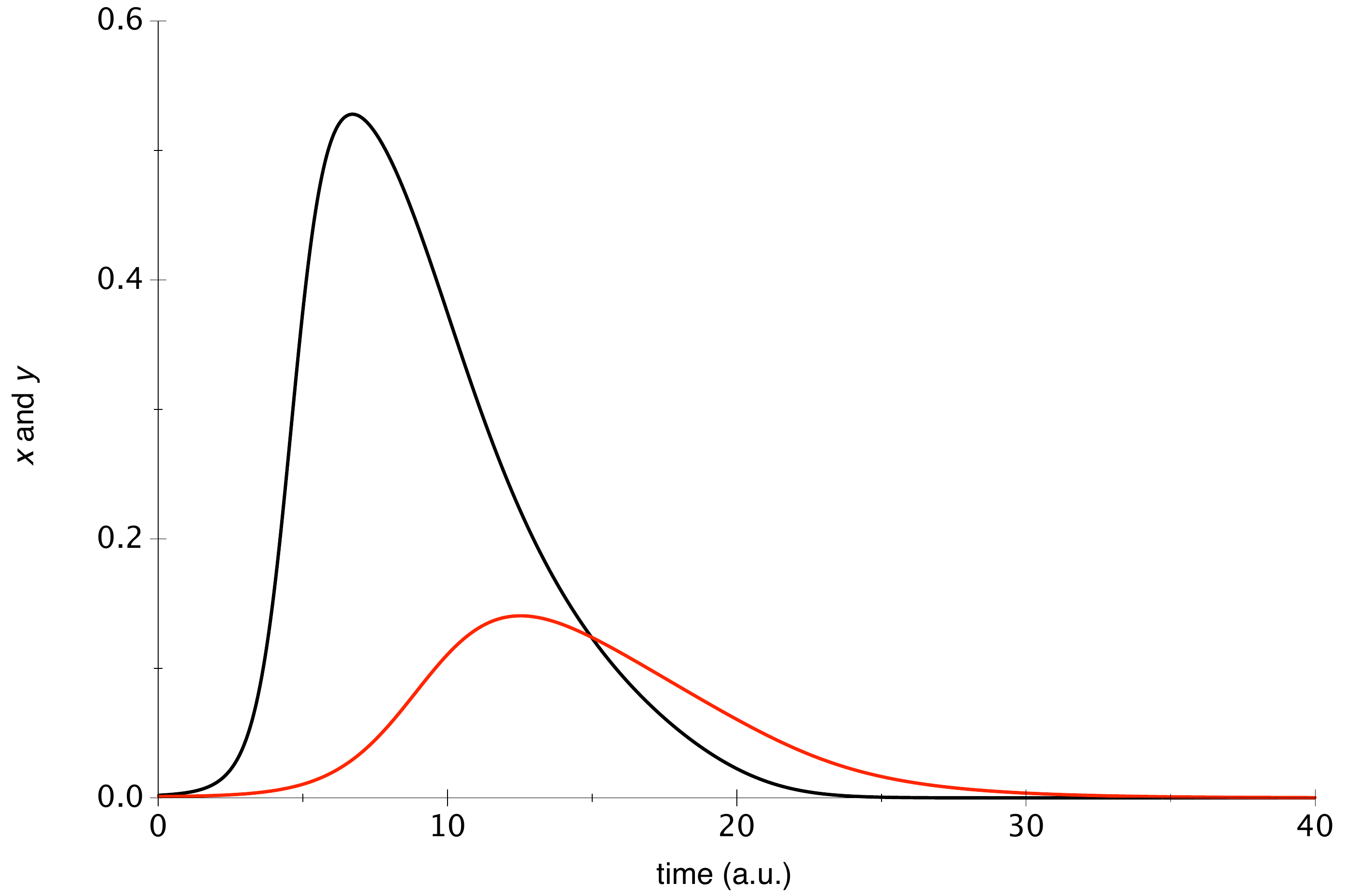}}
\smallskip{\bf Figure \figdef\two}. {\it Evolution of the biomass of two species, versus time, for model (\denn). For the first species (dashed line), $x(0)=2.0~10^{-3}$, $a=3.0$, $f_{xy}=0.1$, $h_{xy}=1.5$, $\lambda=1.2$ and for the second species (full line), $y(0)=1.0~10^{-3}$, $b=1.0$, $f_{yx}=1.0$, $h_{yx}=0.5$, $\mu=1.3$.}
\medskip
Figure \two~shows such an example.

We turn now to the case where at least one of the fixed points is not zero. The fixed points are given by the equations
$$x_*(ax_*^2+ah_{xy} x_*y_*+(\lambda-a)x_*+(h_{xy}\lambda-af_{xy})y_*)=0\eqdaf\dtri$$
$$y_*(by_*^2+bh_{yx} x_*y_*+(\mu-b)y_*+(h_{yx}\mu-bf_{yx})x_*)=0\eqdbf\dtri$$
When one of the fixed points, say $y_*$, is zero the other one is given by $x_*=1-\lambda/a$. The characteristic equation has two roots, one which is by construction negative $\omega_1=-ax_*$, and one $\omega_2=\mu-bf_{yx}/h_{yx}$ which can be made to take a negative value (meaning that the fixed point is attractive) with the appropriate choice of the parameters.
Assuming that both $x_*$ and $y_*$ are non-zero one must solve (\dtri) leading, upon elimination, to a cubic equation for of them. In order to bypass this difficulty we shall use the trick, already introduced in [\handy], solving for two of the parameters which enter linearly (here $\lambda$ and $\mu$), which is tantamount to considering $x_*$ and $y_*$ as new parameters. Once we do this the characteristic equation can be written in a most compact way
$$\displaylines{\omega^2(x_*+h_{xy} y_*)^2(y_*+h_{yx} x_*)^2+  
\omega\Big(a x_*(y_*+h_{yx} x_*)^2\big((x_*+h_{xy} y_*)^2+y_*(f_{xy}-h_{xy})\big)+by_*(x_*+h_{xy} y_*)^2\big((y_*+h_{yx} x_*)^2\hfill\cr\hfill+x_*(f_{yx}-h_{yx})\big)\Big)
+abx_*y_*\Big((f_{xy} -h_{xy})y_*(y_*+h_{yx} x_*)^2+(f_{yx}-h_{yx})x_*(x_*+h_{xy} y_*)^2\hfill\cr\hfill+(x_*+h_{xy} y_*)^2(y_*+h_{yx} x_*)^2\Big)=0.\quad\eqdisp\dtes\cr}$$
Given this complicated expression, it is not possible to perform the stability analysis in full generality. However if one fixes the values of the parameters of the problem, checking the stability of the fixed points becomes elementary.

Extending the model to more than two species is straightforward but an important choice is in order.  It is clear that the possibility of every species to interact (either in competition or in cooperation) with all the other species should be present in the model. This means that one must extend the capacity function but this can be done in different ways. Let us illustrate this in the case of three interacting species, $x$, $y$, $z$. Focusing on the capacity of species $x$, two possibilities exist, an ``additive'' one

$${ x+f_{xy} y+f_{xz} z\over x+h_{xy} y+h_{xz} z},$$ 
or a ``multiplicative'' one
$$\left({ x+f_{xy} y\over x+h_{xy} y}\right)\left({ x+f_{xz} z\over x+h_{xz} z}.\right),$$

Extending this to a higher number of species does not present any difficulty, but one must make a choice between an additive or multiplicative interaction for every new species.

Constructing discrete analogues of equations, like (\denn), involving two species, or in fact any higher number of species, is elementary if one follows the prescription introduced for the single species equation (\dtri): one discretises the $x^2$ term as $x_nx_{n+1}$ and  staggers the $-\lambda x$ term, obtaining $-\lambda x_{n+1}$. This guarantees the positivity of the solution of the system, once one starts with positive initial conditions and positive coefficients. 

\bigskip
4. {\scap Results}
\medskip
In this section we are going to present results obtained through simulations based on our model (\denn). We shall start by considering two species. In order to be able to refer to each of them in a simple way we shall call them A and B.


The first results we are going to present are meant to highlight the
effect of competition among species. In the first simulation we focus
on a case where species B is not influenced by A but
presents a hindrance to the latter. 
\medskip
\centerline{\includegraphics[width=10cm,keepaspectratio]{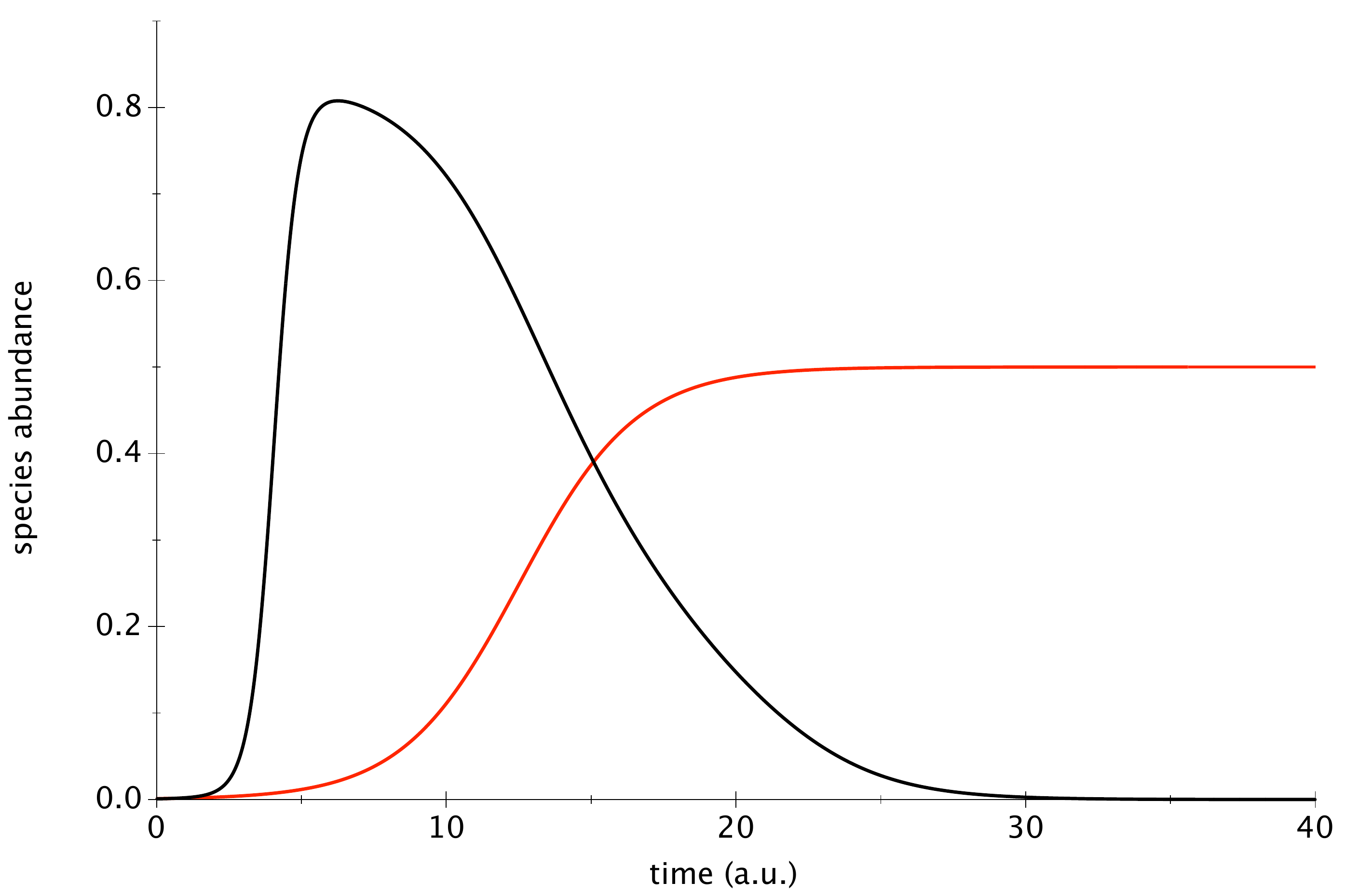}}
\smallskip{\bf Figure \figdef\three}. {\it Evolution of the abundance of the two species $A$ and $B$ versus time, with only B-towards-A hindrance. For species A (black line), $x(0)=8.0~10^{-4}$, $a=3.0$, $f_{xy}=0.0$, $h_{xy}=1.0$, $\lambda=0.5$ and for species B (red line), $y(0)=1.0~10^{-3}$, $b=1.0$, $f_{yx}=0.0$, $h_{yx}=0.0$, $\mu=0.5$.}
\medskip

The result of this competition, (shown in Figure \three), is that species A grows during a certain time
but starts declining when B initiates its growth, and, ultimately,
disappears. The scenario illustrates the principle of species
exclusion [\Gause1935].

Next we show a few more example of the effect of competition among species. 
\medskip
\centerline{\includegraphics[width=10cm,keepaspectratio]{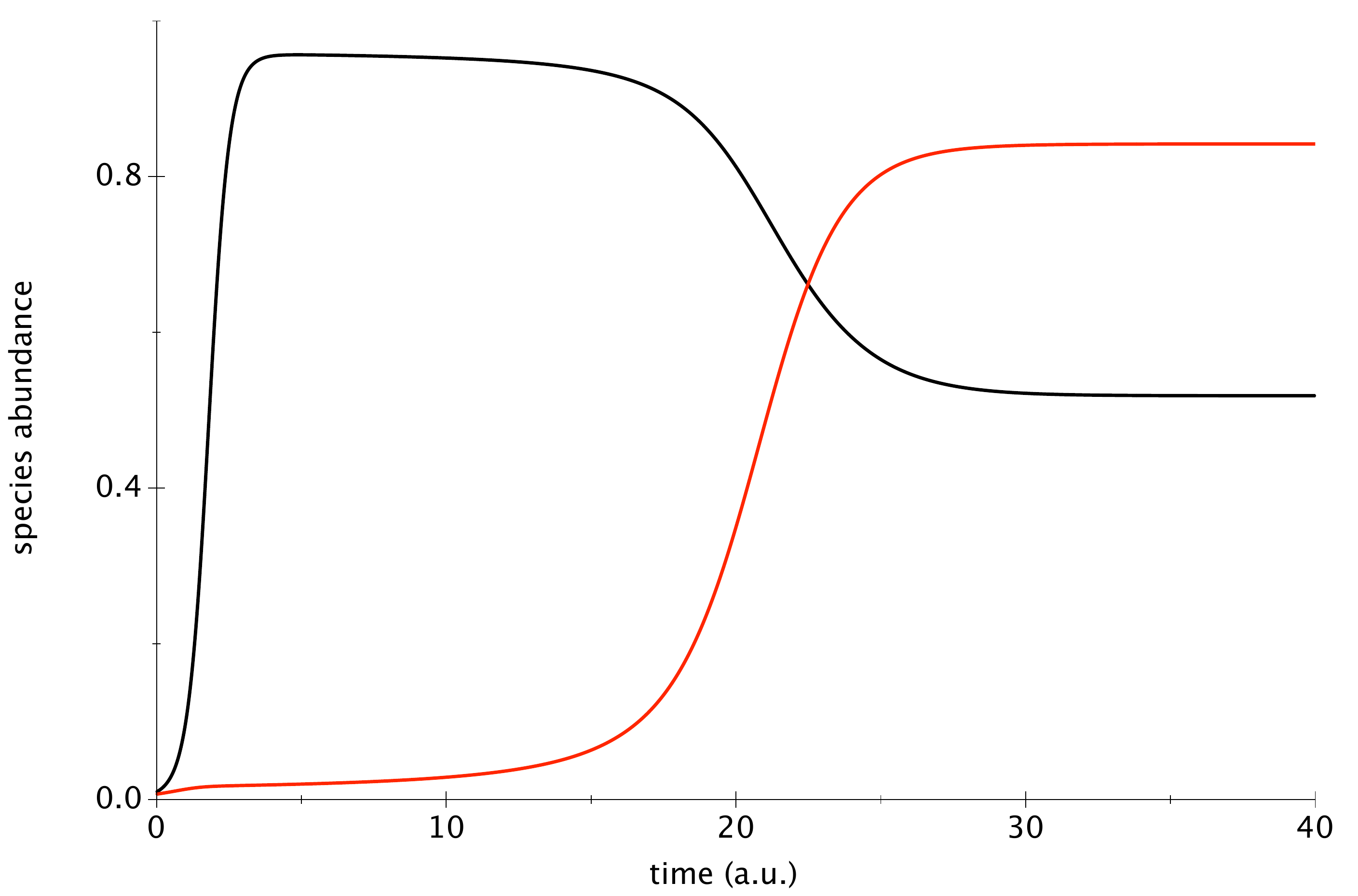}}
\smallskip{\bf Figure \figdef\four}. {\it Evolution of the abundance of the two species $A$ and $B$ versus time, with B-towards-A and A-towards-B hindrance. For species A (black line), $x(0)=1.0~10^{-2}$, $a=3.0$, $f_{xy}=0.0$, $h_{xy}=0.5$, $\lambda=0.1$ and for species B (red line), $y(0)=7.0~10^{-3}$, $b=1.0$, $f_{yx}=0.0$, $h_{yx}=0.1$, $\mu=0.1$.}
\medskip
Figure \four~depicts a situation similar to that of \three~
albeit one where there exist a mutual, but rather mild, competition
between the two species. Species A has now the time to reach a
temporary equilibrium, but the growth of species B disturbs it
pushing the population of species A towards a new, lower, fixed
point. In this case, both species can coexist at the cost of
reduced fitness.

Another interesting effect, resulting from competition between species
is the one known as the `founder' effect also known as priority effect
[\Demeester2016], in which new invasions depend on the order of
previous colonization. 
\medskip
\centerline{\includegraphics[width=10cm,keepaspectratio]{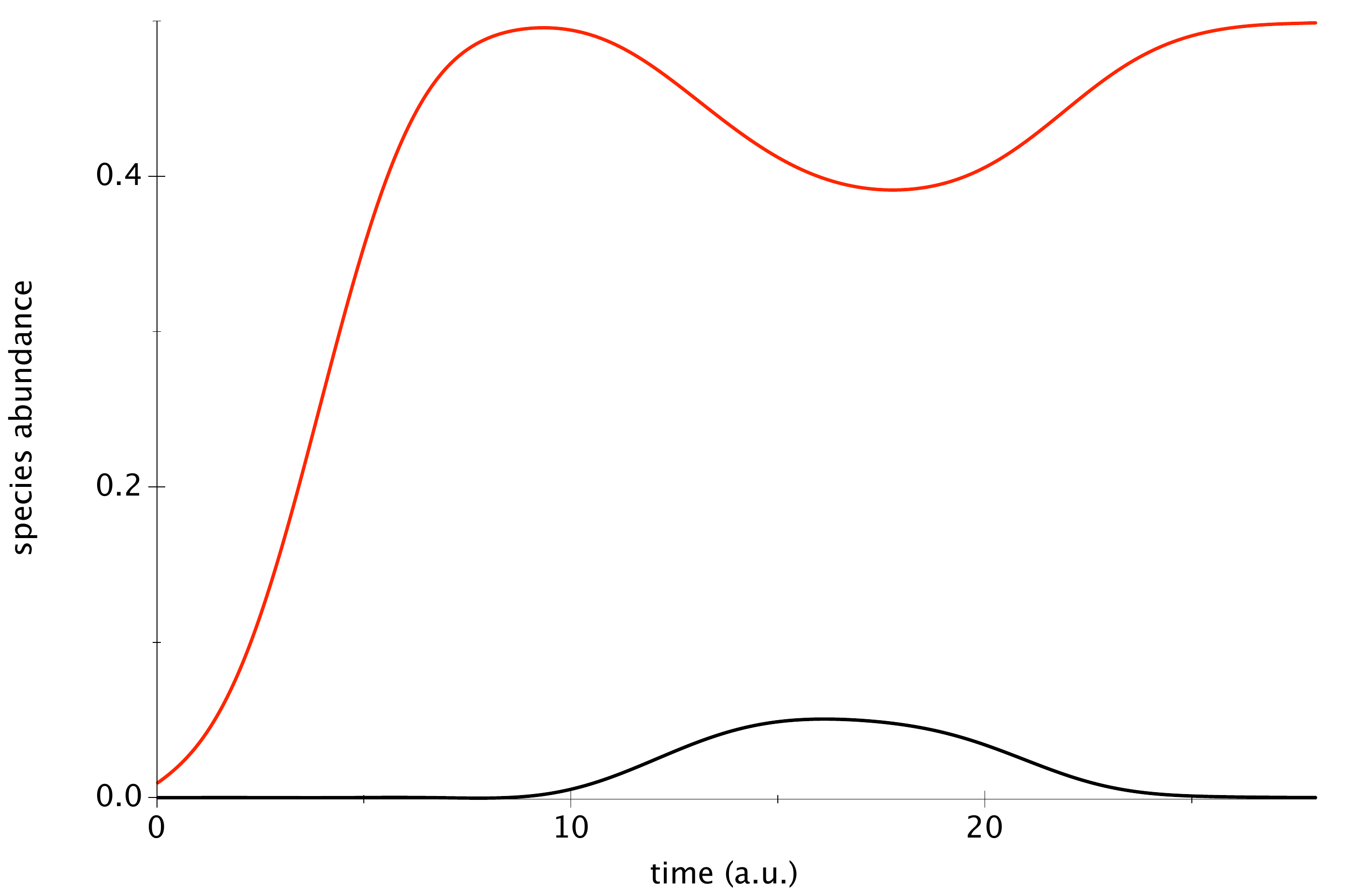}}
\smallskip{\bf Figure \figdef\five}. {\it Evolution of the abundance of the two species $A$ and $B$ versus time. For species A (black line), $a=15.4$, $f_{xy}=0.0$, $h_{xy}=1.0$, $\lambda=1.0$ and for species B (red line), $y(0)=1.0~10^{-2}$, $b=2.0$, $f_{yx}=0.0$, $h_{yx}=1.0$, $\mu=1.0$. The species A is introduced at time $t=10$ with an exponential function $x(t)=0.11(1-\exp(-0.6(t-10)/500))$.}
\medskip
In Figure \five~we present a simulation of this effect: species B being well entrenched hinders the growth of A, through competition, and forces it to disappearance, paying the price with only a temporary decrease of its population.
However it must be pointed out that by fine tuning the amount of competition between the two species we can have a situation similar to the one present in Figure \three, whereupon species A does not totally disappear but reaches some finite equilibrium.
\medskip
\centerline{\includegraphics[width=10cm,keepaspectratio]{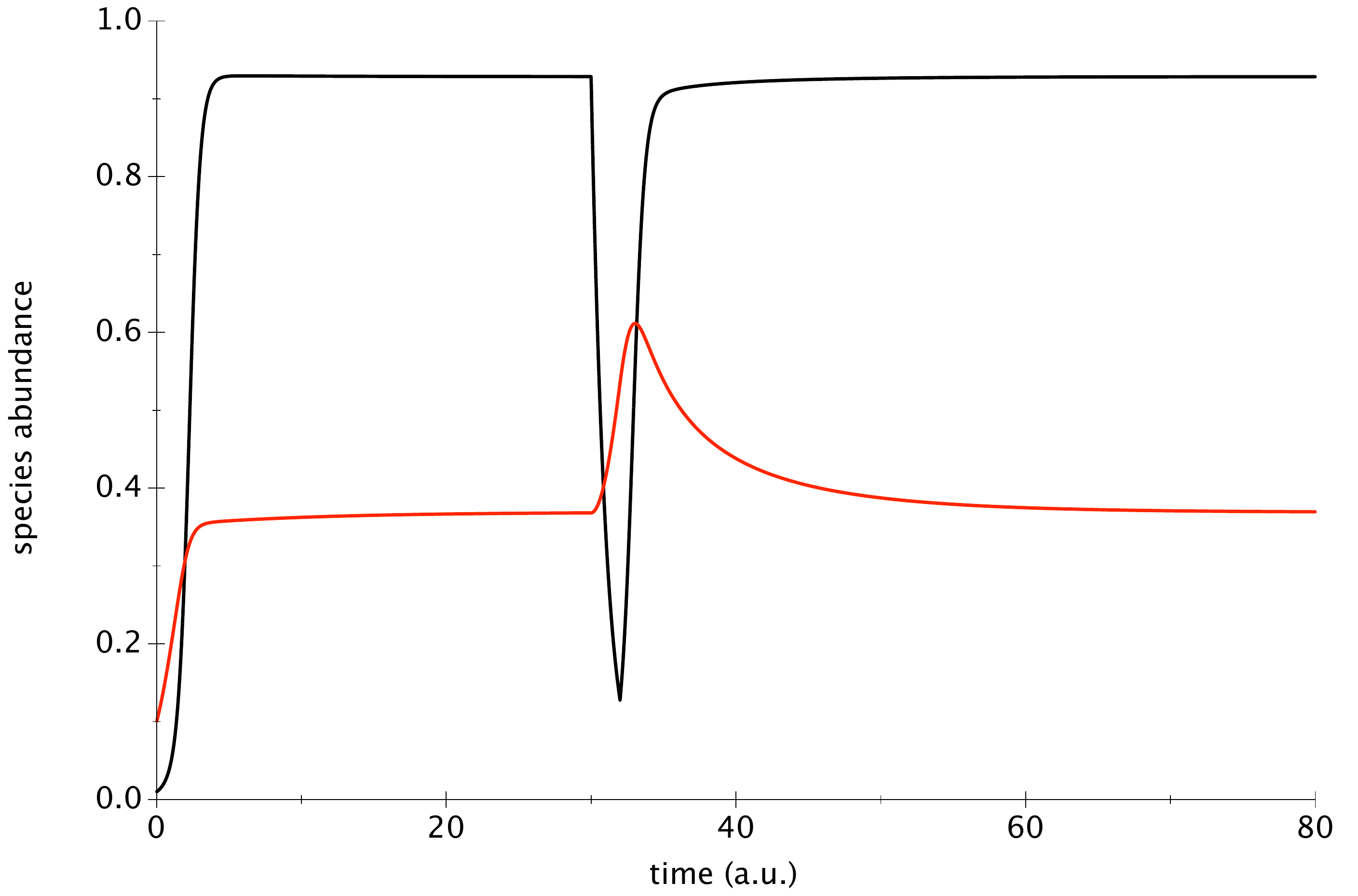}}
\smallskip{\bf Figure \figdef\six}. {\it Evolution of the abundance of two species versus time, and the effect of a tornado. For species A (black line), $x(0)=1.0~10^{-2}$, $a=3.0$, $f_{xy}=0.0$, $h_{xy}=0.1$, $\lambda=0.1$ and for species B (red line), $y(0)=0.1$, $b=1.0$, $f_{yx}=0$, $h_{yx}=0.45$, $\mu=0.1$. For $t \geq 30$ and $t \leq 32$, $x(t)=x(30)\exp(-2(t-30)/2))$.}
\medskip

The final competition-induced example of evolution is the one we call the 'tornado effect', see Figure \six. The two species coexist but A is more abundant than B. Then due to an extreme event (hence the tornado moniker) the population of A is decimated. During the time where the headcount of A is low species B profits in order to increase its population. However this respite is temporary, and once the population of A starts growing the population of B diminishes reverting to the previous equilibrium.
It is interesting to point out here that if by chance species B
was the one to be more abundant before the devastation the subsequent
evolution would have been the same i.e. species A would
temporarily grow only to revert to the previous equilibrium after a
short lapse of time. The restoration towards the previous steady
state is in agreement with empirical data observed in natural forests
after the passage of hurricanes [\refdef{\Woods2009}] (see Appendix~A).

\medskip
\centerline{\includegraphics[width=10cm,keepaspectratio]{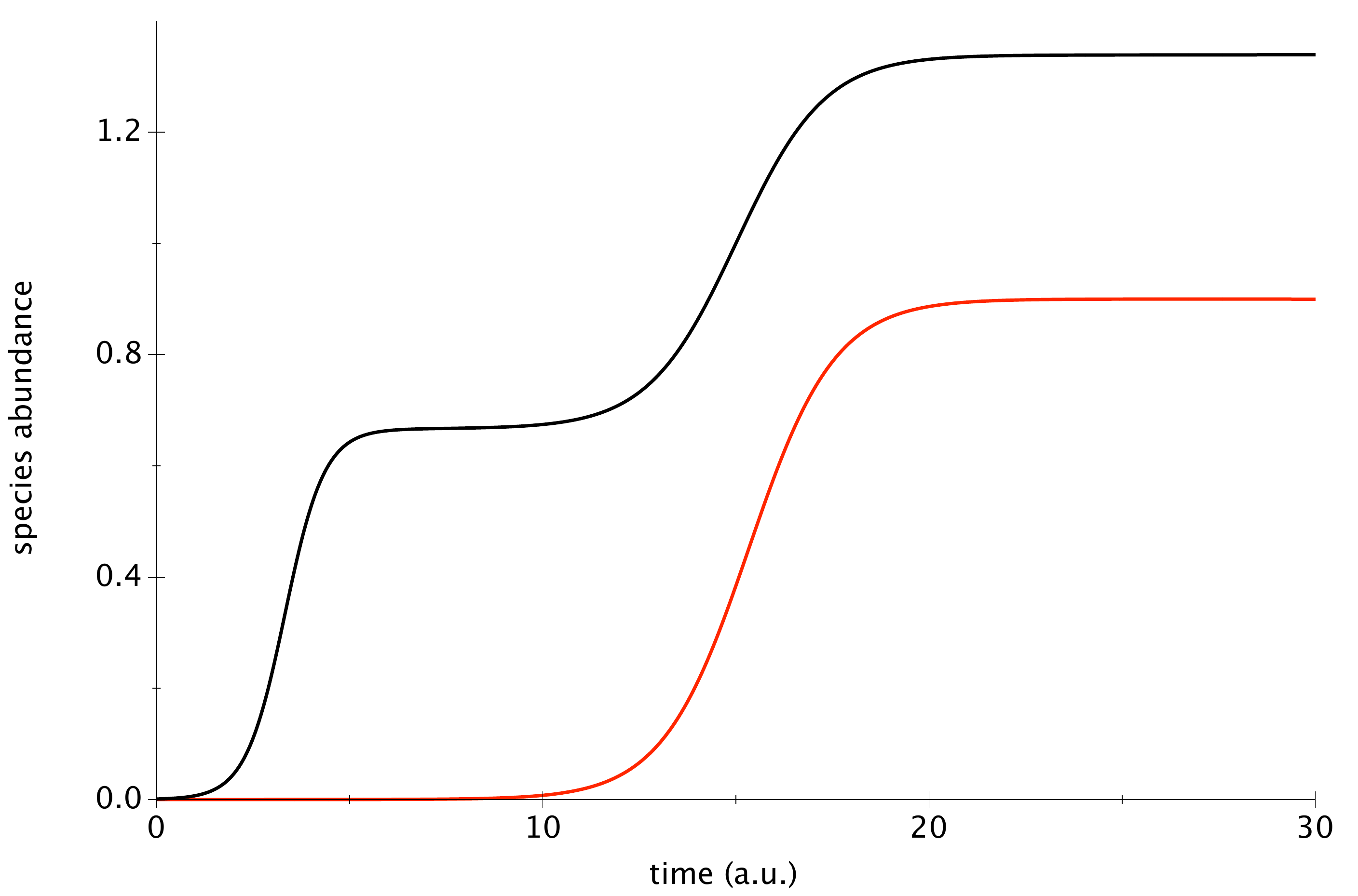}}
\smallskip{\bf Figure \figdef\seven}. {\it Evolution of the abundance of two species versus time, with competition and cooperation. For species A (black line), $x(0)=1.0~10^{-3}$, $a=3.0$, $f_{xy}=1.0$, $h_{xy}=0.0$, $\lambda=1.0$ and for species B (red line) $y(0)=1.0~10^{-6}$, $b=1.0$, $f_{yx}=0.0$, $h_{yx}=0.0$, $\mu=0.1$.}
\medskip

We now turn to the case where cooperation is present among the
species. Figure \seven~shows what happens in such a case. Species A
becomes well established and reaches a first equilibrium
point. Species B grows out of a smaller initial population an thus
needs some time in order to reach a substantial headcount, it 
facilitates the growth of A pushing it to a higher equilibrium
point. 
It is interesting at this point to wonder what would remain of the
founder effect were species to exhibit some degree of cooperation.   
\medskip
\centerline{\includegraphics[width=10cm,keepaspectratio]{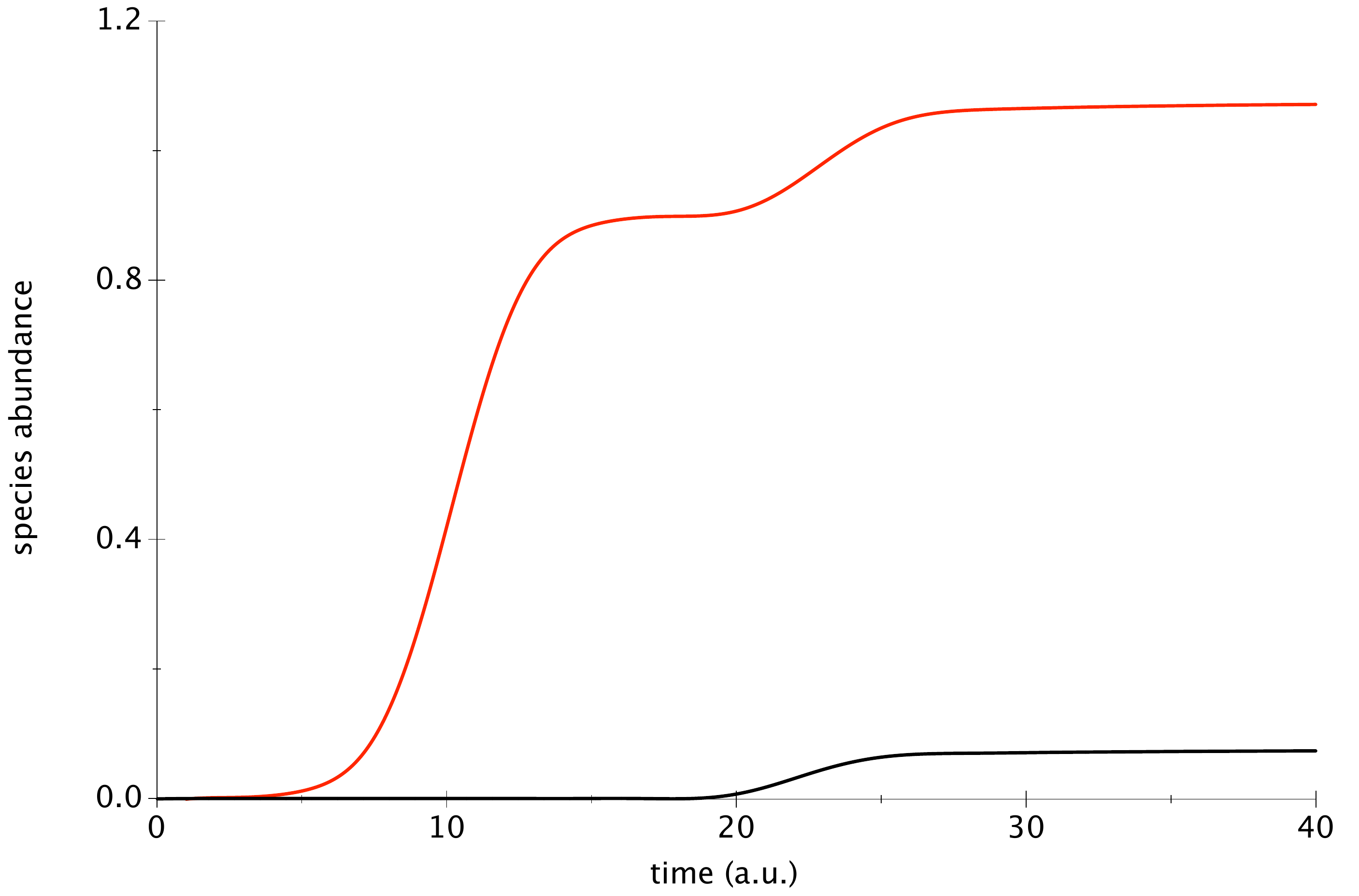}}
\smallskip{\bf Figure \figdef\eight}. {\it Evolution of the abundance of the two species $A$ and $B$ versus time, with competition and cooperation, in the context of the founder effect. For species A (black line), $a=3.0$, $f_{xy}=0.1$, $h_{xy}=1.5$, $\lambda=0.1$ and for species B (red line) $y(0)=1.0~10^{-4}$, $b=1.0$, $f_{yx}=2.5$, $h_{yx}=0.0$, $\mu=0.1$. Species A is introduced at time $t=10$ with an exponential function $x(t)=0.11(1-\exp(-0.6(t-10)/500))$.}
\medskip

In Figure \eight~we present the results of a simulation where the late arriving species A can persist indefinitely thanks to a, loosely adjusted, rate of cooperation with the already entrenched B. It is remarkable that the appearance of species A facilitates the growth of B pushing it to a higher headcount. 

In fact, depending on the amount of cooperation between the two
species,  the late appearing A can grow up to an equilibrium point
surpassing that of B. Notably, this result describes the arrival
of rare and specialized species in mature ecosystems such as mature
soils communities [\refdef{\Coleman2018}] or above ground species in primary
tropical forests [\refdef{\Raven}]. These species can only explore
specific niches which are created and maintained by the continuous
ecological activity of other species. Hence, the establishment of
these rare groups  depends strictly on the presence of a previously
established community providing the necessary supporting conditions.

The final example involving two species is somewhat special in the sense that our two different species could be two specimens of the same one. 
It is well known that the interactions between different species but also between one specimen of one species with its neighbour can be modulated by the external stress. It has been shown that these interactions should shift from negative (competition) to positive (facilitation) across gradients of increasing external stress. This effect has been called the stress gradient hypothesis.
To measure the interactions, researchers usually perform neighbour removal experiments, that consist in removing the neighbours around a target plant. After some time of growth of the target plant, its biomass is measured and compared to the biomass when the neighbours are left intact. If the biomass without neighbours is lower than with intact neighbours, this means that there was facilitation between the target plant and its neighbours. A biomass without neighbours larger than with intact neighbours signals competition. 
For example, in [\refdef{\Namazi2017}], nurse trees create a favorable environment in a very arid region, shedding the sun light and keeping the moisture of the soil. The authors showed that the outcome of the interactions between understory species is competition at the center of the canopy, where the stress is low, and switch to facilitation at the border of the canopy where the stress is higher.

We model the neighbour removal experiments by  running the simulations with two species that represent two groups of the same species (in that case they have the same growth and death rates) or different species. Then the neighbours are removed and their concentration is set to zero. The neighbours are called here species A and the target plant is called species B.
\medskip
\centerline{\includegraphics[width=10cm,keepaspectratio]{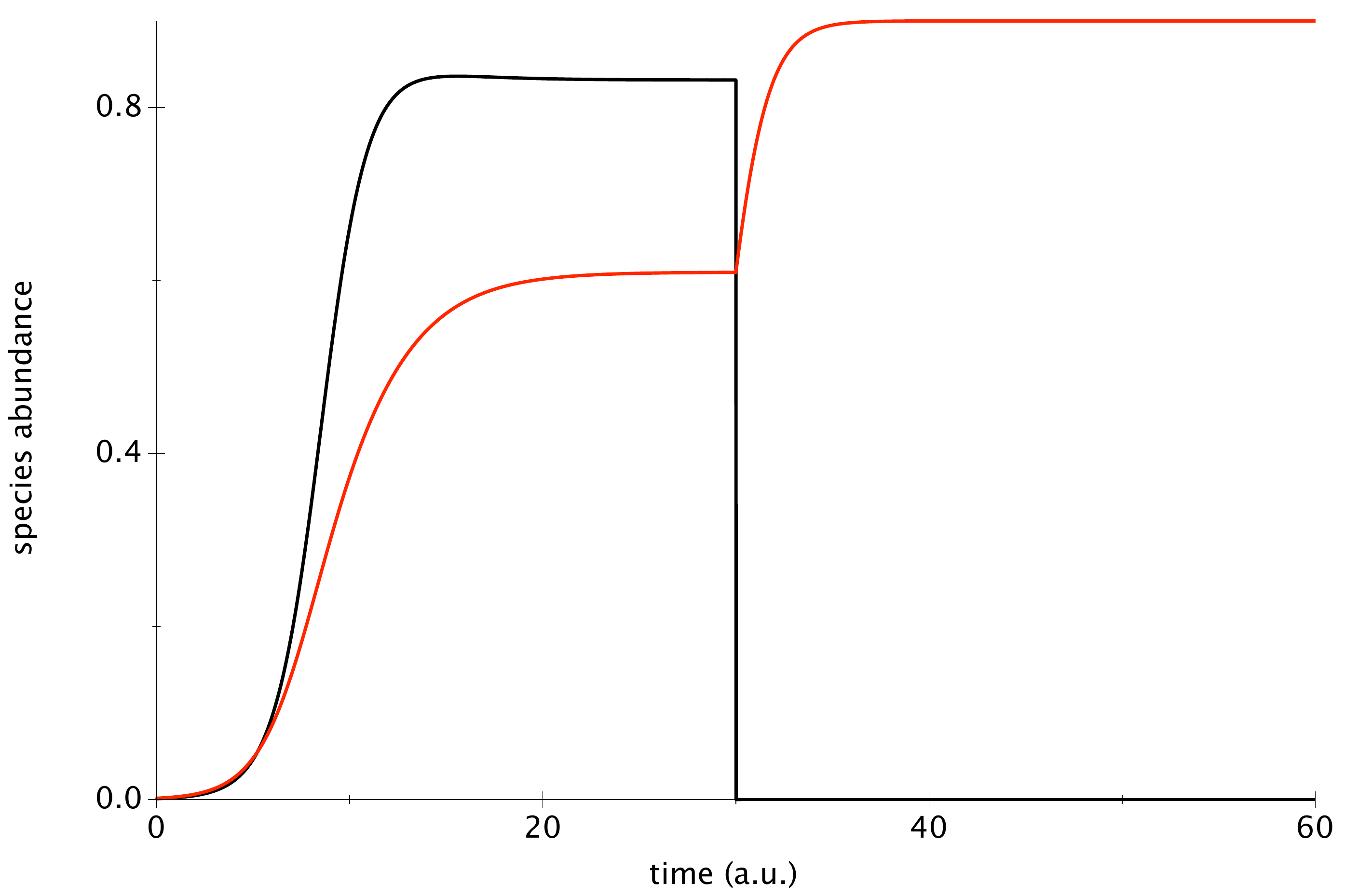}}
\smallskip{\bf Figure \figdef\nine}. {\it Evolution of the abundance of the two species, $A$ and $B$ versus time, with competition. For species A (black line), $x(0)=1.0~10^{-3}$, $a=1.0$, $f_{xy}=0.0$, $h_{xy}=0.1$, $\lambda=0.1$ and for species B (red line), $y(0)=1.5~10^{-3}$, $b=1.0$, $f_{yx}=0.0$, $h_{yx}=0.3$, $\mu=0.1$. The abundance of species A, $x$, is set to zero at time $t=30$.}
\medskip

With pure competition between the two groups, the target species (B) starts to grow when the neighbours (A) are removed (Figure \nine).

\medskip
\centerline{\includegraphics[width=10cm,keepaspectratio]{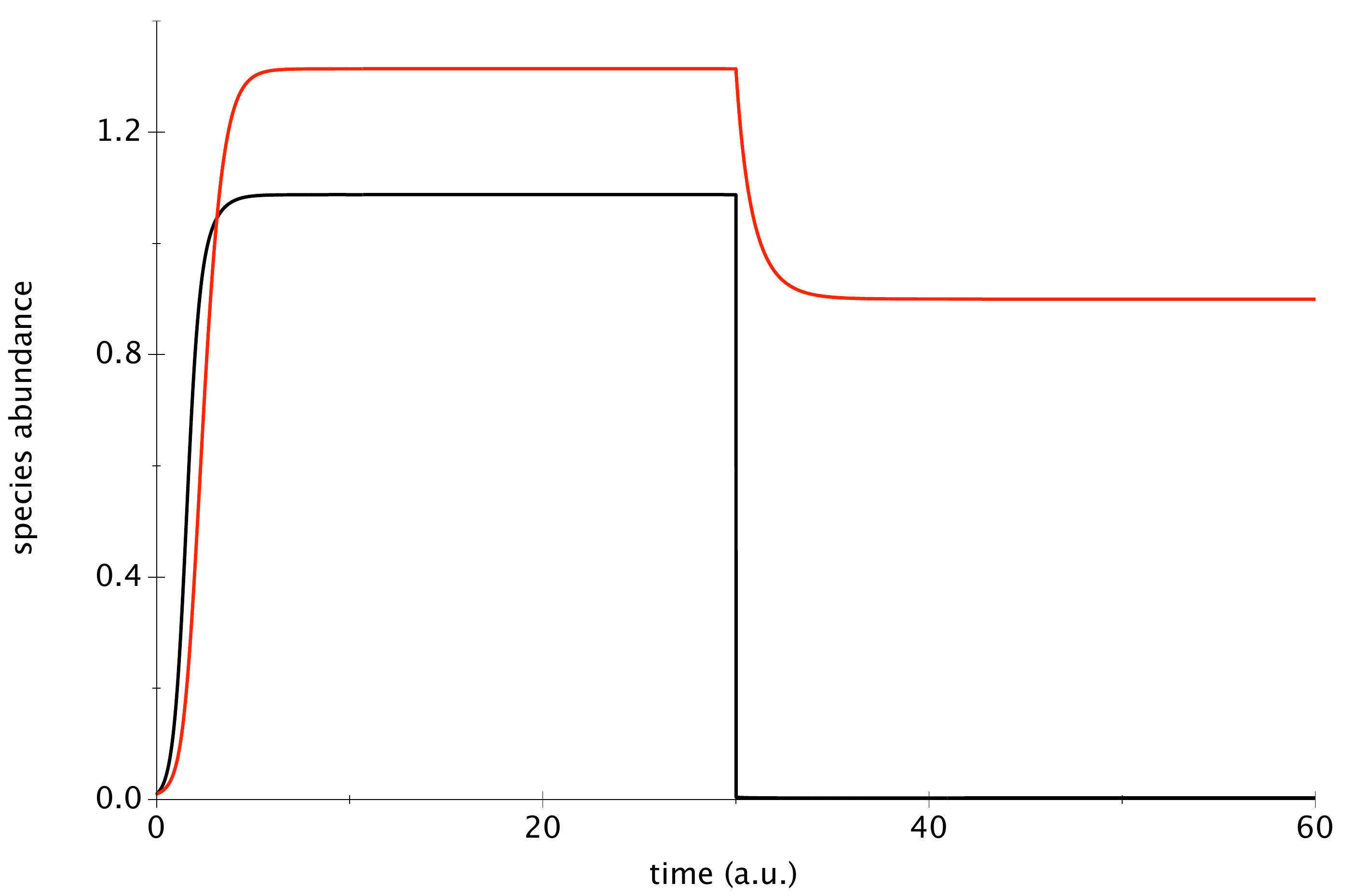}}
\smallskip{\bf Figure \figdef\ten}. {\it Evolution of the abundance of the two species, $A$ (full line) and $B$ (dashed line) versus time, with facilitation. For species A (black line), $x(0)=1.0~10^{-2}$, $a=3.0$, $f_{xy}=0.1$, $h_{xy}=0.0$, $\lambda=0.1$ and for species B (red line), $y(0)=1.0~10^{-2}$, $b=1.0$, $f_{yx}=0.5$, $h_{yx}=0.0$, $\mu=0.1$. The abundance of species A, $x$, is set to zero at time $t=30$.}
\medskip

With pure facilitation, the abundance of the target species decreases and a new equilibrium is reached, lower than the one before the removal (Figure \ten).  

We turn now to the consideration of three species and their
interactions. The simulations were performed using the `additive'
assumption for the capacity function. While the `multiplicative' one
has the advantage of introducing pairwise interactions it has also a
major drawback: when two species happen to vanish the corresponding
factor in the capacity function becomes indeterminate. The additive
ansatz is not plagued by this and, as a matter of fact, it does
simplify calculations. Since we are going to work with three species we denote them as A, B and C. We keep the same notations for the facilitation and hindrance parameters as for two species. The death {rates} of species A, B and C are respectively $\lambda$, $\mu$ and $\eta$.
We do not intend to present a complete exploration of the parameter space and the possible consequences. We shall rather present a selection of results which highlight the richness of the interactions. 

\medskip
\centerline{\includegraphics[width=10cm,keepaspectratio]{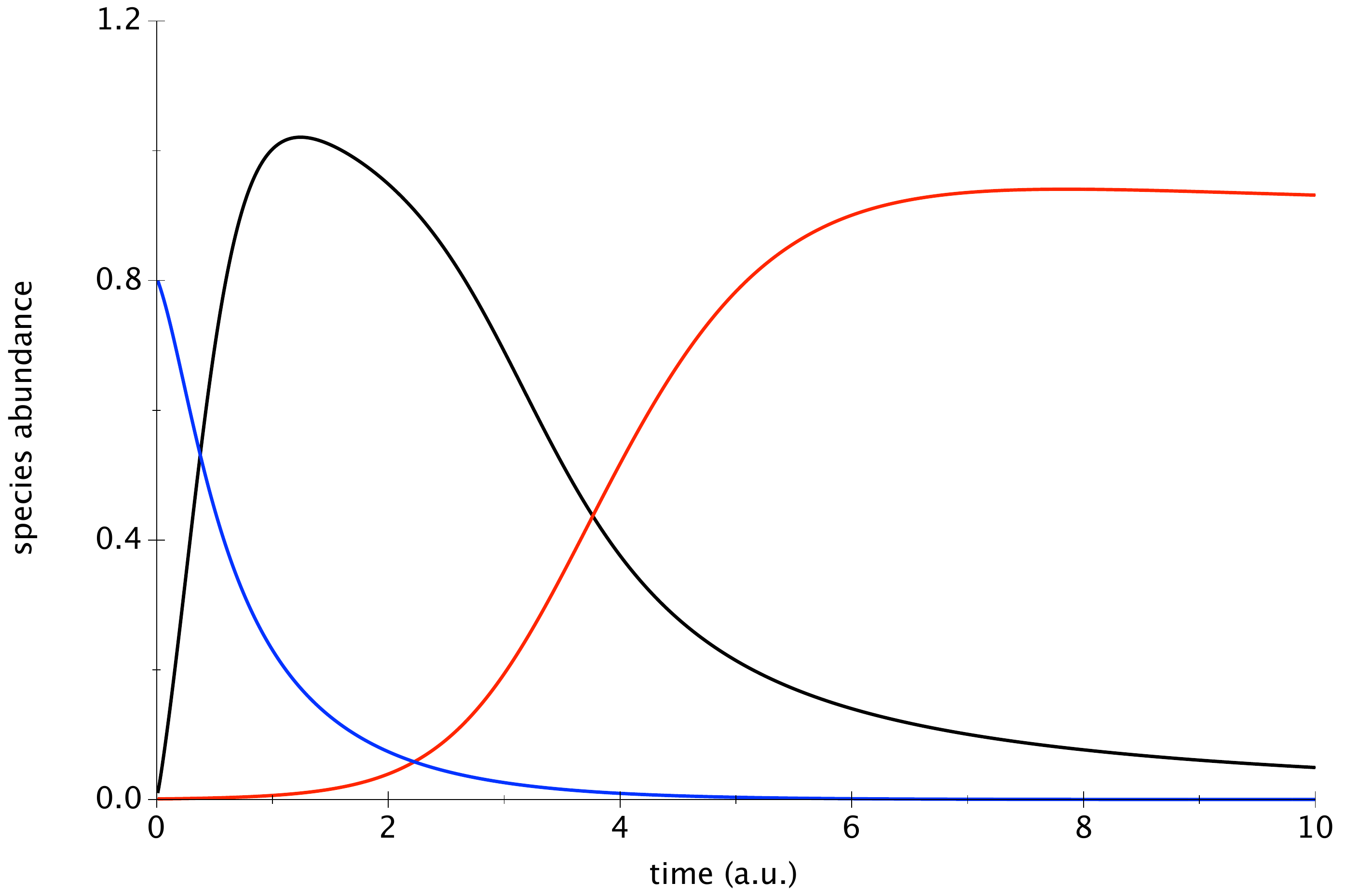}}
\smallskip{\bf Figure \figdef\eleven}. {\it Evolution of the abundance of three species versus time. For species A (black line), $x(0)=1.0~10^{-2}$, $a=3.0$, $\lambda=0.1$, $f_{xy}=0.0$, $f_{xz}=0.5$, $h_{xy}=3.0$, $h_{xz}=0.0$. For species B (red line), $y(0)=1.0~10^{-3}$, $b=1.0$, $\mu=0.1$, $f_{yx}=1.0$, $f_{yz}=0.0$, $h_{yx}=0.5$, $h_{yz}=0.0$. For species C (blue line), $z(0)=0.8$, $c=2.0$, $\eta=1.0$, $f_{zy}=0.0$, $f_{zx}=0.0$, $h_{zy}=2.0$, $h_{zx}=2.0$.}
\medskip

In Figure \eleven~we show an evolution which is reminiscent of the famous
Tilman drawing. In this classic example, Tilman shows the successional
process involving grass and non-grass species in old fields in a more competitive than facilitating case.

\centerline{\includegraphics[width=10cm,keepaspectratio]{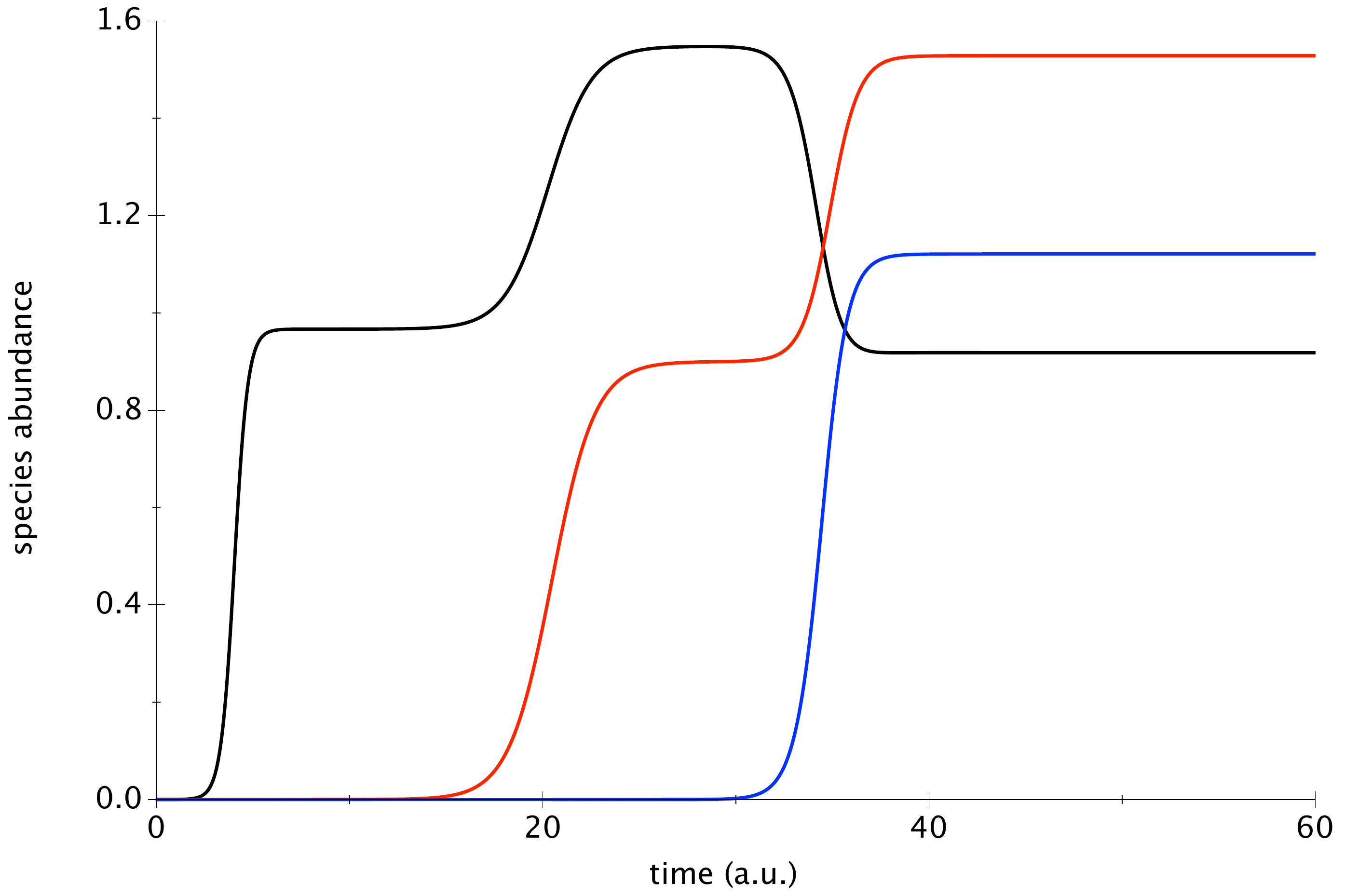}}
\smallskip{\bf Figure \figdef\twelve}. {\it Evolution of the abundance of three species versus time. For species A (black line), $x(0)=1.0~10^{-5}$, $a=3.0$, $\lambda=0.1$, $f_{xy}=1.0$, $f_{xz}=0.5$, $h_{xy}=0.0$, $h_{xz}=2.0$. For species B (red line), $y(0)=1.0~10^{-8}$, $b=1.0$, $\mu=0.1$, $f_{yx}=1.0$, $f_{yz}=3.0$, $h_{yx}=1.0$, $h_{yz}=1.0$. For species C (blue line), $z(0)=1.0~10^{-8}$, $c=2.0$, $\eta=0.1$, $f_{zy}=2.0$, $f_{zx}=0.0$, $h_{zy}=1.0$, $h_{zx}=1.0$.}
\medskip
Figure \twelve~shows an evolution of three species where the first two have a mutual facilitation and evolve in a pattern similar to that of Figure \seven. However when, much later, the third species, C, arrives hindering the growth of species A but facilitating that of B, the equilibria do change and the relative abundances are reshuffled.
\medskip
\centerline{\includegraphics[width=10cm,keepaspectratio]{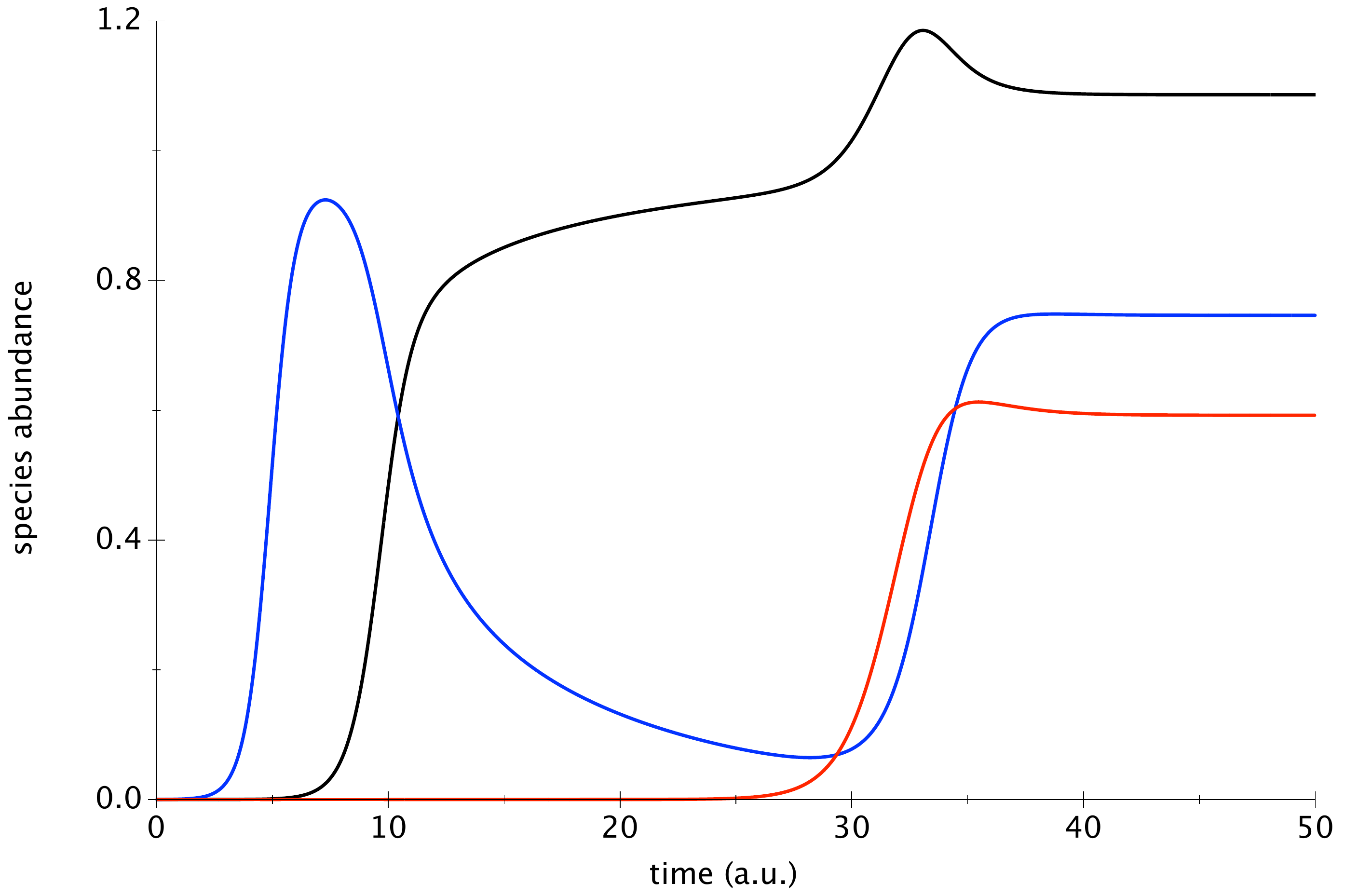}}
\smallskip{\bf Figure \figdef\thirteen}. {\it Evolution of the abundance of three species versus time. For species A (black line), $x(0)=1.0~10^{-6}$, $a=3.0$, $\lambda=0.1$, $f_{xy}=1.0$, $f_{xz}=0.5$, $h_{xy}=0.0$, $h_{xz}=1.0$. For species B (red line), $y(0)=1.0~10^{-7}$, $b=1.0$, $\mu=0.1$, $f_{yx}=1.0$, $f_{yz}=0.0$, $h_{yx}=1.0$, $h_{yz}=1.0$. For species C (blue line), $z(0)=1.0~10^{-4}$, $c=2.0$, $\eta=0.1$, $f_{zy}=2.0$, $f_{zx}=0.0$, $h_{zy}=1.0$, $h_{zx}=1.0$.}
\medskip
An interesting situation is presented in Figure \thirteen. Here species C starts growing but its competition with A pushes it to near extinction. However the late arrival of species B, which is in mutual cooperation with the other two not only revives species C but it also pushes A to a higher equilibrium.

In the figures above we presented situations where two of species
disappeared in the long run (Figure \eleven) and two where all three
species survived (Figures \twelve  and \thirteen). It goes without saying that a situation where only one of the three species arrives in the long run is equally possible.

\bigskip
5. {\scap Discussion}
\medskip

\bigskip
5.1 {\scap The interaction compass in a single model}
\medskip

The presence of mutable interactions in plants explained by the stress
gradient hypothesis underscores the necessity to include positive,
negative, and neutral interactions in biological systems [\Hoek2016].
Our model captures the various types of ecological interaction
(mutualism, neutralism, pure competition, etc) depending on the values
(or intervals) of the parameters $(f_{xy},f_{yx})$ and $(h_{xy},
h_{yx})$ in (19a) and (19b).

The first pair quantifies facilitation while the second one
corresponds to hindrances between the two species. The mathematical
structure of (19), together with the generalization for more species,
ensures the dynamics remains regular at any time instant. In practice,
this means that populations never diverge, even for simultaneous
cooperation. Note that we do not impose explicit resource dynamics,
which is often used to limit the exponential growth in mutualist
partner species. This simplification allow us to explain the complex
patterns that emerge from species interaction alone without prior
knowledge about the organisms' physiology or the details of the
spatial distribution of resources, extraction, exchange, spatial
diffusion and movement patterns of individuals or subgroups when
applicable.

As a result of this compact formulation, the plasticity of ecological
interactions is attributed to the dependence of $\{f_{xy},f_{yx},
h_{xy},h_{yx}\}$ on external stress factors.
{ The interaction is positive, negative, or neutral depending on
the sign of the difference between facilitation and hindrance
parameters, $f_{xy} - h_{xy}$. } 
The case  $f_{xy} = h_{xy} =0$ bring us to the familiar scenario of
independent species but the situation is less intuitive for other
values. In particular, the interspecific competition is higher than
the intraspecific with $f_{xy} = h_{xy} > 1$ but the facilitation
term compensates and produces a null effect in the end. 
In Figures~{\nine} and {\ten}, we simulate experiments
in which the plants surrounding a target species are completely removed
from the environment at $t=t_0$. The discontinuity of the growth of
the target species is proportional to $( h_{xy} - f_{xy} )$. In Figure~{\nine}
 (Figure~\ten), the discontinuity is positive (negative) in agreement with
the competition (cooperation) driven process. The removal of smaller
amounts of neighbours (instead of all of them) corresponds to
discontinuities with different amplitudes. If the fraction of removed
individuals can be estimated, then one can compute the coefficients
$f_{xy}$ and $h_{xy}$ under the assumption that stress factors remain
constant.

\bigskip
5.2 {\scap Asymmetric facilitation improves survival and invasion odds}
\medskip

We do not claim our model is the only one to describe the full
range of interactions in ecology. On the contrary, there exists a
plethora of models that produce the same effect as discussed in the
previous sections. Here, the population maximum for each species is
dynamically adjusted to reflect the influences of other elements of the
community. It expands the foundations put forth by Kuno by considering
the positive role of facilitators in addition to the negative impact
of competitors. The results in Figures~{\three} and {\four} reinforce this
message in which coexistence occurs between competitive species as
long as the intraspecific competition  is larger than the
interspecific one, i.e. $h_{xy} < 1$ and $h_{yx} < 1$, as reported in
[\kuno], and in agreement with the Chesson's coexistence theory
[\Chesson2000].

In the context of a pure competitive scenario, the equilibrium densities 
sit at unitary values after species exclusion, or values lower than
one which reflects the sub-optimal growth under competitive
coexistence. In the case of species exclusion, the fittest species
tend to dominate and exclude the other. The exception for this rule
occurs when the less fit species B excludes the other due to a stronger
competitive barrier for the invader (priority effect) which we depict
in Figure~\five. { The most notable consequences of the ecological
priority effect can be appreciated in the wake of large disturbances
or during the colonization of new regions. In both cases, the effect
can lead ecosystems with similar biological background into distinct
successional routes and thus different community structures.}



A remarkable phenomenon takes place when the facilitation is
asymmetric. In Figure~\six, a species with higher fitness (B) invades an
environment with another population already entrenched. Instead of
triggering the priority effect and thus species exclusion (see
Figure~\five), the species with lower fitness (A) but higher initial
headcount keeps its dominance and, simultaneously, limits the growth
of B. The inversion is only possible because the asymmetric
facilitation pushes up the sub-optimal species A which in turn
increases the growth pressure on the invader B. This symbiosis loosely
resembles the husbandry of species B by A in a cooperative-competitive
process.

Conversely, the successful invasion of A into a region with a stable
population of species B depends, crucially, on biological processes
that mirror facilitation (see Figure~\seven). This happens because the
stability of the population B, as an initial condition, reduces the
window of opportunity for species A to invade since it is less apt.
The facilitation in this context balances the negative effects of
the competition and, for large enough values of facilitation, A can
overthrow the more adapted species B reversing the core idea behind
species exclusion.

\bigskip
5.3 {\scap Applications to conservation efforts}
\medskip


The loss of biodiversity has been acknowledged as a global and
widespread issue that directly or indirectly affects human livelihood
and health. It has also been linked to the { reduced exploration of
ecosystems} and, thus, the elimination of niches which creates a
downward spiral tendency that only accelerates the exclusion of
species. Yet, ecological conservation faces a tremendous number of
challenges. The most pressing ones take place at global scales such as
climate changes and habitat degradation. At local scales, subtle
effects surface ranging from biological invasions by aggressive
competitors to changes in interspecies interaction induced by varying
environmental stress (anthropogenic or not). A question of practical
interest therefore arises on how these effects impact the ecological
succession at typical time scales pertinent to recovery efforts.  

To answer that question, it is convenient first to revisit the famous
experiments put forth by Tilman in the late 80s [\Tilman1990]. In
short, the experiments describe the various stages of  the ecological
succession observed in old fields in Cedar Creek Natural History Area
(CCNHA), Minnesota. Old fields comprise   the collection of abandoned
agricultural areas sorted according to their abandonment date, where
each area corresponds to a certain stage of the succession. Data on
plant diversity, biomass, and abundance span more than 80 years and
remains to date one of the most iconic representations of plant
succession. The experiments conducted at CCNHA have served to identify
the environmental constraints and the basic mechanisms associated with
plant interactions in order to build the foundations of a modern
predictive ecological theory.

In the early stages, the successional
structure at CCNHA depicts the alternate dominance between short-lived
agricultural weeds and perennial grass with notable
differences in the trait distribution for colonization, resource
capture, and biomass distribution. For the three more common and
representative species, the plant succession occurs between the {\it
  Agrostis scabra}, followed by {\it Agropyron repens}, and finally by
the superior competitor {\it Schizachyrium scoparium}. The former
displays enhanced colonization traits with improved seed production
and dispersal. As a result, {\it Agrostis} tends to be the early
colonizer,later displaced by {\it Agropyron}, which allocates more
biomass to the quick formation of roots and specialized root tissue
(rhizome) to increase nitrogen capture. {\it Schizachyrium} lacks the
competitive ability to be an early colonizer or compete with the fast
growing {\it Agrostis}. However, once {\it Agropyron} becomes
dominant, {\it Schizachyrium} profits from its lower nitrogen
requirements to persist in the habitat to become the dominant
species. In Figure~\eleven, the generalization of our model with three
competing species recovers the general aspects of the data
presented by Tilman.

At early stages of plant succession, the general trend mirrors the
exclusion of species. However, a growing body of work suggests that
maximum biodiversity takes place at mid-succession stages, as reported
for boreal forests [\refdef{\Taylor2020}]. The strong resource
competition experienced by fast growing colonizers is replaced, in
part, by a mixture of hindrance and facilitation between
shade-tolerant species, leading to the creation of new niches and the
efficient partition of existing ones. While this can be complicated in
competition-driven models, it is only a matter of parameter adjustment
in our model. In fact, it is easy to tune the parameters for three
species in order to produce periods with reduced growth, resembling
the plateau-like curves generally associated with plant succession. 
Similar to the case of two species, facilitation parameters favor
species coexistence and they can even invert the final community
structure depending on their relative amplitude compared to hindrance
ones. Figure~{\twelve} exhibits a case in which a sub-optimal species become
the dominant in a given environment by profiting from the growth of an
invader. 
Note that if one is not aware of the presence of the invader,
one would (incorrectly) conclude that the subdominant species somehow
evolved or improved its competitive abilities in a very short
timeframe. 

The biodiversity of a given region is also linked to the presence of
foundation species, functional groups of habit-forming species that
mold the ecosystem [\Bruno2003]. Their decline often precedes the
passage towards a new stage of the ecological succession, which can be
either an advancement or a regression. The latter can occur by natural
means (natural fires and similar events) or by degradation of the
habitat. In the first case, the population of foundation species
experiences a reduction but rebounds later (see Figure~\seven), while in the
second case the foundation species become less efficient
competitors. For mature ecosystems, the regression to previous
successional stages implies a loss in biodiversity which might take
years to decades to recover from. Therefore, foundation
species are generally accepted as the primary targets for conservation
efforts.  

However, the restoration of degraded habitats still remains a
challenge and often requires long-term commitment and resources. In
the short-term, another burdensome task concerns the restoration of
ill-adapted organisms in degraded conditions: the target species for
conservation actions might disappear due to the pressure of
competitors more adapted to the current conditions. Thus, it is
reasonable to look for mechanisms that mitigate the abundance loss of 
foundation species if one hopes to maintain the correct functioning of
the system. A pathway to reverse the decline consists of actions which
re-introduce or stimulate the growth of supporting species. Figure~{\thirteen}
depicts the facilitation-driven mechanism to preserve the foundation
species. The introduction of the supporting species compensates the
reduced competitiveness of the target ones and reverses the population
exclusion. Because the supporting species are also successful invaders,
they thrive in the degraded habitat and can maintain themselves
without external help. In practice, the early introduction of
supporting species just accelerate the return to the target stage of the
ecological succession. Without them, the return interval would be
increased by the waiting time necessary for the system to meet the
required conditions for the successful invasions of foundation
species, in agreement with the emergence of facilitating processes in
mid-succession stages [\Taylor2020].

\bigskip
6. {\scap Conclusion}
\medskip

In this paper, we have proposed a new model that extends the Kuno
model to address multispecies population dynamics focusing on the
facilitation and cooperative regime.
Our extension implicitly includes the stress gradient hypothesis, i.e. facilitation parameters.
It allows the model to address non-trivial transients, such as biological invasion, ecological succession, Tilman drawings etc. 
Also, it naturally avoids divergences always presenting finite asymptotic behaviors, i.e., the species carrying capacity is an emergent quantity. 
Other than allowing robust behaviors, the model also allows fine tuning, which can be an effective mechanism in the recuperation of ecological systems. 
The main issue we want to stress is the richness of transient behavior, with support of ecological examples. 
The long time behavior can be analytically inferred by stability analysis.
Although we have not performed an exhaustive study in the parameter space, which we have kept to the minimum dimensions, we are able to retrieve several known ecological regimes.
Nevertheless, a more systematic analysis of the parameter space could reveal other interesting regimes especially when the number of species is increased.

\bigskip
{\scap Acknowledgements}
\medskip
{\it Funding.} This work was supported by CNRS, CNPq 309851/2018-1,
and CAPES 88887.475028/2020-00.

\medskip
{\it Declaration of interest.} The Authors declare that they have no
known competing financial interests or personal relationships that
could have appeared to influence the work reported in this paper.

\bigskip
{\scap References}
\medskip

\begin{enumerate}
\item[]{[\Hoek2016]} Tim A. Hoek, Kevin Axelrod, Tommaso Biancalani, Eugene A. Yurtsev, Jinghui Liu, and Jeff Gore, ``Resource availability modulates the cooperative and competitive nature of a microbial cross-feeding mutualism,'' PLOS Biology 14, 1-17 (2016).

\item[]{[\Begon2021]} Michael Begon and Colin R. Townsend, ``Ecology:
From Individuals to Ecosystems,''  Wiley (2021). 
\item[]{[\Hastings2007]} Alan Hastings, James Byers, Jeff Crooks, Kim Cuddington, Clive Jones, John Lambrinos, Theresa Talley, and William Wilson, ``Ecosystem engineering in space and time,'' Ecology letters 10, 153-64 (2007).
\item[]{[\Ellner2018]} Stephen P. Ellner, Robin E. Snyder, Peter B. Adler, and Giles Hooker, ``An expanded modern coexistence theory for empirical applications,'' Ecology Letters 22, 3-18 (2019).
\item[]{[\Miele2019]} Vincent Miele, Christian Guill, Rodrigo Ramos-Jiliberto, and Sonia K\'{e}fi, ``Non-trophic interactions strengthen the diversity—functioning relationship in an ecological bioenergetic network model,'' PLOS Computational Biology 15, 1-20 (2019).

\item[]{[\volterra]} V. Volterra, ``Variazione e fluttuazione del numero d'individui in specie animali conviventi,'' Mem. Accad. Nazionale Lincei 2, 31-113 (1926).

\item[]{[\mcarthur]}  R. MacArthur, and R. Levins, ``Competition, habitat selection, and character displacement in a patchy environment,'' Nat. Acad. Sci. USA 51, 1207-1210 (1964).

\item[]{[\levin]} Simon A. Levin, ``Community Equilibria and Stability, and an Extension of the Competitive Exclusion Principle,'' American Naturalist  104(939), 413-423 (1970)

\item[]{[\Gause1935]} G.F. Gause and A.A Witt, ``Behavior of mixed populations and the problem of natural selection,'' The American Naturalist 69, 596-609 (1935).

\item[]{[\Hutchinson1957]} G.E. Hutchinson, ``Concluding Remarks,'' Cold Spring Harbor Symposia on Quantitative Biology 22, 415-427 (1957)

\item[]{[\Fischer2019]} Dylan G. Fischer, Joseph A. Antos, Abir Biswas, and Donald B. Zobel, ``Understorey succession after burial by tephra from mount st. helens,'' Journal of Ecology 107, 531-544 (2019).
\item[]{[\Clark2019]} Adam Thomas Clark, Johannes M. H. Knops, and Dave Tilman, ``Contingent factors explain average divergence in functional composition over 88 years of old field succession,'' Journal of Ecology 107, 545-558 (2019).
\item[]{[\Prach2019]} Karel Prach and Lawrence R. Walker, ``Differences between primary and secondary plant succession among biomes of the world,'' Journal of Ecology 107, 510-516 (2019). 
\item[]{[\Changb2019]} Cynthia C. Chang, Charles B. Halpern, Joseph A. Antos, Meghan L. Avolio, Abir Biswas, James E. Cook, Roger del Moral, Dylan G. Fischer, Andrés Holz, Robert J. Pabst, Mark E. Swanson, and Donald B. Zobel, ``Testing conceptual models of early plant succession across a disturbance gradient,'' Journal of Ecology 107, 517-530 (2019).
\item[]{[\Hutchinsonb1961]}  G.E. Hutchinson, ``The paradox of the plankton,'' The American Naturalist 95, 137-145 (1961).
\item[]{[\Palmer1994]} Michael W. Palmer, ``Variation in species richness: Towards a unification of hypotheses,'' Folia Geobotanica et Phytotaxonomica 29, 511 (1994).
\item[]{[\Chesson2000]} Peter Chesson, ``Mechanisms of maintenance of species diversity,'' Annual Review of Ecology and Systematics 31, 343-366 (2000).
\item[]{[\Yuyama2018]} Ikuko Yuyama, Masakazu Ishikawa, Masafumi Nozawa, Masa-aki Yoshida, and Kazuho Ikeo, ``Transcriptomic changes with increasing algal symbiont reveal the detailed process underlying establishment of coral-algal symbiosis,'' Scientific Reports 8, 16802 (2018).
\item[]{[\Fisher2019]} Rebecca Fisher, Pia Bessell-Browne, and Ross Jones, ``Synergistic and antagonistic impacts of suspended sediments and thermal stress on corals,'' Nature Communications 10, 2346 (2019).
\item[]{[\Moyano2020]} Jaime Moyano, Mariano A. Rodriguez-Cabal, and Martin A. Nun\~{e}z, ``Highly invasive tree species are more dependent on mutualisms,'' Ecology 101, e02997 (2020).

\item[]{[\Gross2008]} Kevin Gross, ``Positive interactions among competitors can produce species-rich communities,'' Ecology Letters 11, 929-936 (2008).
\item[]{[\Tilman1990]} David Tilman, ``Constraints and tradeoffs: Toward a predictive theory of competition and succession,'' Oikos 58, 3-15 (1990).
\item[]{[\Muller2018]} Torben Hilmers, Nicolas Friess, Claus B\"{a}ssler, Marco Heurich, Roland Brandl, Hans Pretzsch, Rupert Seidl, and J\"{o}rg  M\"{u}ller, ``Biodiversity along temperate forest succession,'' Journal of Applied Ecology 55, 2756-2766 (2018).
\item[]{[\Bruno2003]} John F. Bruno, John J. Stachowicz, and Mark D. Bertness, ``Inclusion of facilitation into ecological theory,'' Trends in Ecology \& Evolution 18, 119-125 (2003).
\item[]{[\Brunob2016]} Fabio Bulleri, John F. Bruno, Brian R. Silliman, and John J. Stachowicz, ``Facilitation and the niche: implications for coexistence, range shifts and ecosystem functioning,'' Functional Ecology 30, 70-78 (2016).
\item[]{[\Clements1916]} Carnegie Institution of Washington, Carnegie Institution of Washington publication, Vol. no.242 (1916) (Washington,Carnegie Institution of Washington, 1916) p. 658.
\item[]{[\Cramer2008]} Viki Cramer, Richard Hobbs, and Rachel Standish, ``What’s new about old fields? land abandonment and ecosystem assembly,'' Trends in ecology \& evolution 23, 104-12 (2008).
\item[]{[\Estrada2020]} Sergio Estrada-Villegas, Mario Bailon, Jefferson S. Hall, Stefan A. Schnitzer, Benjamin L. Turner, Trevor Caughlin, and Michiel van Breugel, ``Edaphic factors and initial conditions influence successional trajectories of early regenerating tropical dry forests,'' Journal of Ecology 108, 160-174 (2020).
\item[]{[\Peltzer2010]} Duane A. Peltzer, David A. Wardle, Victoria J. Allison, W. Troy Baisden, Richard D. Bardgett, Oliver A. Chadwick, Leo M. Condron, Roger L. Parfitt, Stephen Porder, Sarah J. Richardson, Benjamin L. Turner, Peter M. Vitousek, Joe Walker, and Lawrence R. Walker, ``Understanding ecosystem retrogression,'' Ecological Monographs 80, 509-529 (2010).
\item[]{[\Damschen2008]} Ellen I. Damschen, Lars A. Brudvig, Nick M. Haddad, Douglas J. Levey, John L. Orrock, and Joshua J. Tewksbury, ``The movement ecology and dynamics of plant communities in fragmented landscapes,'' Proceedings of the National Academy of Sciences 105, 19078-19083 (2008).
\item[]{[\Meloni2015]} Fernando Meloni and Elenice M.Varanda, ``Litter and soil arthropod colonization in reforested semi-deciduous seasonal atlantic forests,'' Restoration Ecology 23, 690-697 (2015).

\item[]{[\Demeester2016]} Luc De Meester, Joost Vanoverbeke, Laurens J. Kilsdonk, and Mark C. Urban, ``Evolving perspectives on monopolization and priority effects,'' Trends in Ecology \& Evolution 31, 136-146 (2016).
\item[]{[\Fukami2015]} Tadashi Fukami, ``Historical contingency in community assembly: Integrating niches, species pools, and priority effects,'' Annual Review of Ecology, Evolution, and Systematics 46, 1-23 (2015).
\item[]{[\Fukami2015]} Cynthia C. Chang and Benjamin L. Turner, ``Ecological succession in a changing world,'' Journal of Ecology 107, 503-509 (2019).
\item[]{[\Diagne2021]} Christophe Diagne, Boris Leroy, Anne-Charlotte Vaissiere, Rodolphe E. Gozlan, David Roiz, Ivan Jaric,Jean-Michel Salles,Corey J. A. Bradshaw, and Franck Courchamp, ''Highand rising economic costs of biological invasions worldwide,'' Nature 592, 571-576 (2021).

\item[]{[\malthus]} T. Malthus, ``An Essay on the Principle of Population,'' London: Johnson, (1798).

\item[]{[\verhulst]} P-F. Verhulst, ``Notice sur la loi que la
population suit dans son accroissement,'' Correspondance mathématique et physique 10, 113-121 (1838).

\item[]{[\morisita]} M. Morisita, ``The fitting of the logistic equation
to the rate of increase of population density,'' Population Ecology 7, 52-55 (1965)

\item[]{[\lotka]} A. J. Lotka, ``Contribution to quantitative parasitology,'' J. Wash. Acad. Sci. 13, 152-158 (1923).

\item[]{[\salomon]} M.E. Solomon, ``The Natural Control of Animal Populations,'' J. Anim. Ecology 18, 1 (1949).

\item[]{[\holling]} C.S. Holling, ``Some Characteristics of Simple Types
of Predation and Parasitism'', Can. Entomol. 91,  293 (1959).

\item[]{[\menten]} L. Menten and M. I. Michaelis, ``Die Kinetik der Invertinwirkung,'' Biochem. Z. 49, 333 (1913)

\item[]{[\crypto]} R. Willox, A. Ramani, and B. Grammaticos,  ``A discrete-time model for cryptic oscillations in predator–prey systems,'' Physica D: Nonlinear Phenomena 238(22), 2238-2245 (2009).

\item[]{[\kuno]} E. Kuno, ``Competitive exclusion through reproductive interference,'' Researches on Population Ecology 34(2), 275-284 (1992).

\item[]{[\ramani]} B. Grammaticos, A. Ramani, J. Satsuma, and R. Willox, ``Discretising the Painlevé equations \`{a} la Hirota-Mickens,'' J. Math. Phys. 53, (2012). 

\item[]{[\mickens]} R. E. Mickens, ``Exact Solutions to a Finite-Difference Model of a Nonlinear Reaction-Advection
Equation: Implications for Numerical Analysis,'' Numer. Methods
Partial Differ. Equ. 5(4), 313–325 (1989).

\item[]{[\handy]} B. Grammaticos, R. Willox, and J. Satsuma, ``Revisiting the Human and Nature Dynamics Model,'' Regular and Chaotic Dynamics 25, 178-198 (2020).

 
\item[]{[\Woods2009]} K. D. Woods, ``Multi-decade, spatially explicit
population studies of canopy dynamics in Michigan oldgrowth forests,''
Ecology 90(12), 3587 (2009).

\item[]{[\Coleman2018]} D. C. Coleman, M. A. Callaham, and D. A. Crossley Jr, ``Fundamentals of soil ecology,'' Academic press (2017).

\item[]{[\Raven]} P. H. Raven, R. F. Evert, and S. E. Eichhorn,
``Biology of plants,'' W. H. Freeman (2005).

\item[]{[\Namazi2017]} A. A. Al-Namazi, M. I. El-Bana, and S. P. Bonser,
``Competition and facilitation structure plant communities under nurse
tree canopies in extremely stressful environments,'' Ecology and Evolution  7, 2747-2755 (2017).

\item[][{\Taylor2020}] Anthony R. Taylor, Bilei Gao, Han Y.H. Chen, ``The effect of species diversity on tree growth varies during forest succession in the boreal forest of central Canada,'' Forest Ecology and Management 455, (2020).
\end{enumerate}

\bigskip
{\scap Appendix A}
\medskip

The example explores long-term data of vegetation succession and
recovery after perturbations in a permanent parcel maintained by Woods
[{\Woods2009}], at Dukes Research Natural Area (DRNA), Michigan,
USA. The original data represent the diameter at breast height (DBH) of wood stems discriminated
by plant species in several plots, totaling $20$ ha. The data regard a
chronosequence of $19$ years, from $1989$ to $2007$, as previous
samples were taken at very large intervals. The region is periodically
affected by winds, with a strong event in 2002, highlighting the
relevance of this database to understand recovering processes. 

We explored the DNAR data as a single site, summing the respective DBH
of species by year. Of the $32$ species observed in the region (one is
unknown), $24$ were observed at least once since $1989$. The dimension
of the dataset is reduced by grouping species as follows. We
consider the population DBH over time as only variable and group
species by their relative variation over time. This sorting scheme
eliminates the necessity of ad hoc information about individual species as
per usual in the formal functional analysis. 

More specifically, we employed the Principal Component Analysis (PCA)
based on covariances of species DBH. The PCA indicated at least $4$
relevant dimensions to explain the data dispersion (eigenvalues larger
than $1$) following Kaiser criterion. Next, we calculate the
dissimilarity matrix based on Euclidean distances in species
coordinates. The hierarchical clustering (Ward method) is subsequently
used and produces two groups (Figure~A\figappdef{\aa}), which are for our purposes
considered as distinct functional groups. 

Note that our sorting approach creates competing groups in a simpler
manner compared to traditional trait-driven approaches, which often
require in-depth knowledge about the pertinent species involved. This
is only possible because we only intend to group species based on
their relative dominance. Hence, we denote ``Group 1'' the group
formed by dominant species; and ``Group 2'' the species that grow when
the density of Group 1 decreases. Next, we compute the total DBH per
group as it provides a quantitative functional response, allowing us
to investigate the interactions between both groups in terms of
population dynamics.

Figure~A\figappdef{\ab} depicts the DBH for each group over time. Species in Group
1 are periodically disturbed by strong winds, an external factor that
negatively affects their DBH. The decrease of the dominant group
triggers a brief spurt of growth by species in Group 2, suggesting
a competitive context. Shortly after the disturbances, the population
returns to the original levels, suggesting that the secondary
succession follows similar trajectories in the region.

\medskip
\centerline{\includegraphics[width=10cm,keepaspectratio]{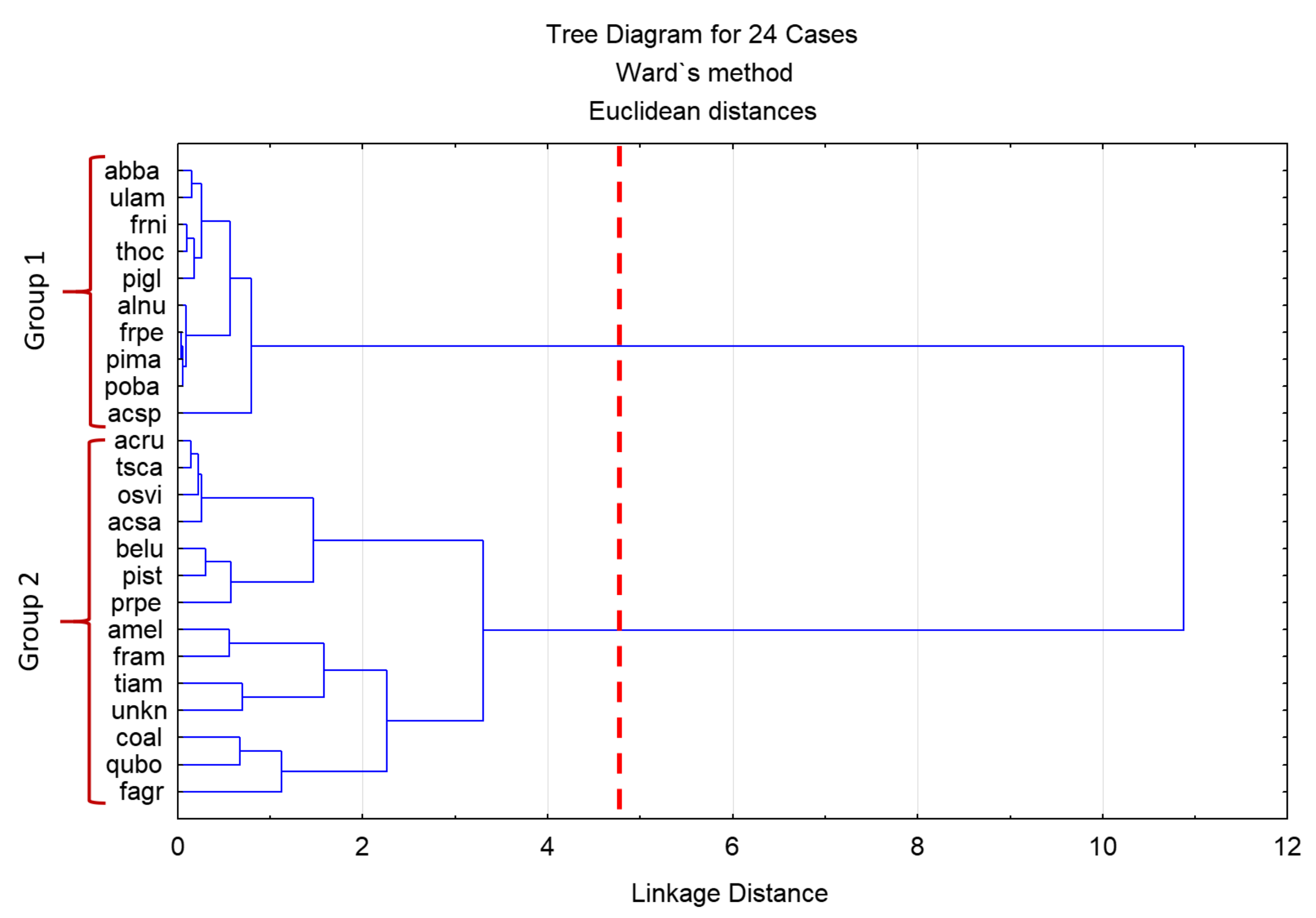}}
Figure~{A.\aa}:
Hierarchical spectral clustering based on species DBH in a
chronosequence. The dashed line represents the cutoff distance linkage
to create two groups. The classification does not rely on species
traits.
\medskip

\medskip
\centerline{\includegraphics[width=15cm,keepaspectratio]{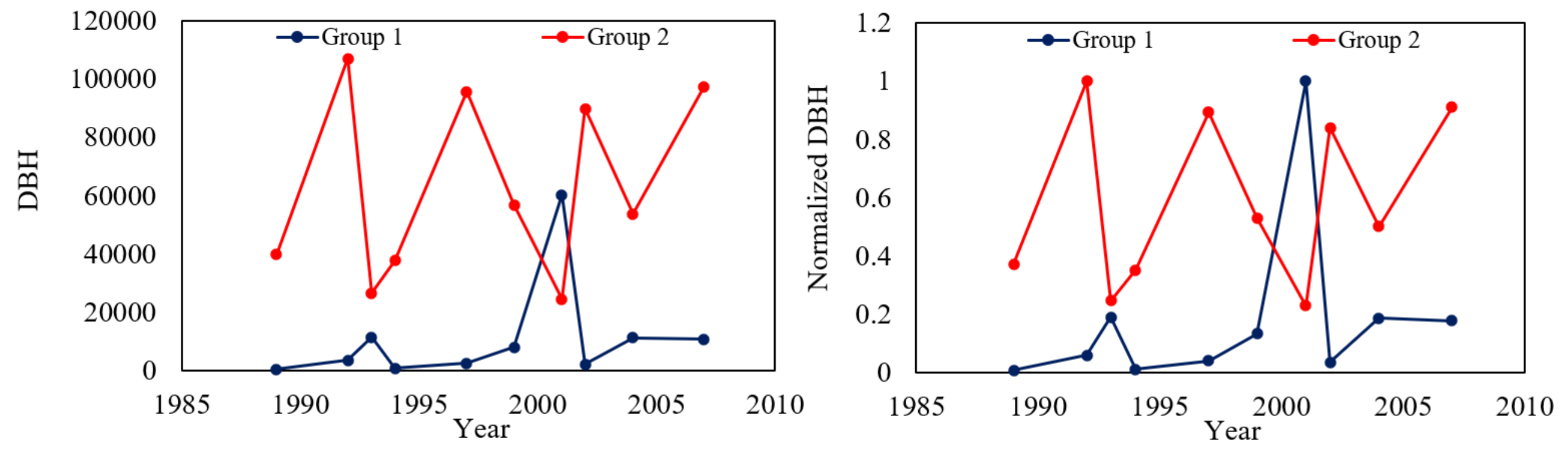}}
Figure~{A.\ab}: Successional dynamics in Dukes Research Natural Area,
Michigan, USA. The curves represent the DBH of tree species after
grouping. (left) Raw DBH values for Group 1 and Group 2 per
year. (right) Normalized values. 
\medskip

\end{document}